\documentclass[aip,pof,amsmath,preprint]{revtex4-1}

\usepackage{graphicx}
\usepackage{color}

\newcommand{\D} {\displaystyle}
\newcommand{\bi}{\boldsymbol}

\newcommand{\unit}[2]{ \ensuremath{{#1} \, \mathrm{#2}}}

\begin{document}

\title{Instabilities and spin-up behaviour of a rotating magnetic field driven
flow in a rectangular cavity}

\author{V. Galindo}
\affiliation{Helmholtz-Zentrum Dresden-Rossendorf, Department Magnetohydrodynamics,
01314 Dresden, Germany}
\author{R. Nauber}
\affiliation{Technische Universit\"at Dresden, Laboratory of Measurement and Sensor
System Technique, 01062 Dresden, Germany}
\author{D. R\"abiger}
\author{S. Franke}
\affiliation{Helmholtz-Zentrum Dresden-Rossendorf, Department Magnetohydrodynamics,
01314 Dresden, Germany}
\author{H. Beyer}
\author{L. B\"uttner}
\author{J. Czarske}
\affiliation{Technische Universit\"at Dresden, Laboratory of Measurement and Sensor
System Technique, 01062 Dresden, Germany}
\author{S. Eckert}
\affiliation{Helmholtz-Zentrum Dresden-Rossendorf, Department Magnetohydrodynamics,
01314 Dresden, Germany}

\date{\today}

\begin{abstract}
This study presents numerical simulations and experiments considering the flow
of an electrically conducting fluid inside a cube driven by a rotating magnetic
field (RMF). The investigations are focused on the spin-up, where
a liquid metal (GaInSn) is suddenly exposed to an azimuthal body force
generated by the RMF, and the subsequent flow development.
The numerical simulations rely on a semi-analytical expression for 
the induced electromagnetic force density in an electrically conducting medium
inside a cuboid container with insulating walls.
Velocity distributions in two perpendicular planes are measured using a novel
dual-plane, two-component ultrasound array Doppler velocimeter (UADV) with 
continuous data streaming, enabling long term measurements for investigating
transient flows. This approach allows to identify the
main emerging flow modes during the transition from a stable to unstable
flow regimes with exponentially growing velocity oscillations using the 
Proper Orthogonal Decomposition (POD) method.

Characteristic frequencies in the oscillating flow regimes are determined
in the super critical range above the critical magnetic
Taylor number $Ta_c \approx 1.26 \times 10^5$, where the transition from the
steady double vortex structure of the secondary flow to an unstable regime 
with exponentially growing oscillations is detected. 

The mean flow structures and the temporal evolution of the flow predicted by the
numerical simulations and observed in experiments are in very good agreement. 
 
\end{abstract}

\maketitle

\section{Introduction}

Electromagnetic flow control is an important and efficient tool for optimizing
fluid flow and transport processes in crystal growth, metallurgy
and metal casting. Various types of tailored AC magnetic fields are applied for
electromagnetic stirring providing a variable and contact-less access to electrically
conducting melts (see Gerbeth et al. \cite{Gerbeth2013} and references therein).
The specific requirements arising from the particular metallurgical or casting 
operation are manifold. For instance, the electromagnetic stirring should provide 
an efficient mixing of the melt or counterbalance buoyancy-driven flows. 
Different types of magnetic fields (as rotating, travelling, pulsating and 
combinations of these three) are available, whereas each field type gives rise to a
more or less distinctive flow pattern.
A deep understanding of the fluid flow properties under the effect of AC magnetic
fields is required for defining optimal configurations and parameters for the
magnetic system.
Within this study we consider the case of rotary stirring in a square cross-section,
for which the process of billet or bloom casting in steel production can serve as
a prominent example from industrial practice.
Fabricating multicrystalline silicon for photovoltaic modules often involves a
rectangular crucible cross section and time-varying magnetic fields from resistive heaters.
A great deal of work has been done yet for investigating
the properties of fluid flow driven by a rotating magnetic field (RMF) inside a
cylindrical vessel. For a detailed review about RMF-driven flows we refer the 
reader to the classical paper of Davidson and Hunt \cite{davidsonhunt1987} 
or Priede and Gelfgat \cite{gelfgatpriede1995}. The transition from a steady to
a time-dependent flow regime was studied among others by Barz et al. \cite{barz1997},
Kaiser and Benz \cite{kaiser1998} and Witkowski et al. \cite{Wikowski1999}.
The authors tried to find a critical non-dimensional magnetic Taylor number for
the onset of the instabilities using different methods for direct numerical 
simulation. Recently, Grants and Gerbeth \cite{Grants2001,Grants2002,Grants2003}
conclude that the RMF-driven flows in a cylinder become unstable first to 
non-axisymmetric,
azimuthally periodic perturbations at diameter-to-height aspect ratios $AR$ 
between $0.5$ and $2$. Ungarish \cite{Ungarish1997} and Nikrityuk 
et al. \cite{Niki2005} considered the so-called spin-up process of a developing
flow when the fluid at rest is exposed to a suddenly applied RMF. The numerical
simulations were confirmed experimentally by flow measurements performed by
R\"abiger et al. \cite{Raebiger2010}. Evolving perturbations of the double vortex
structure just above the instability threshold occur in form of 
Taylor--G\"ortler vortices. The rotational symmetry of the flow structure is 
kept while a first Taylor--G\"ortler vortex pair has been formed as closed 
rings along the cylinder perimeter. The transition to a three-dimensional
flow in the side layers occurs by advection, precession and splitting of the
Taylor - G\"ortler vortex rings \cite{vogt2012eif}.

In contrast, the number of studies dealing with rotary stirring in square or
rectangular cross sections appears to be rather small.  Dubke et 
al. \cite{dubke1988a, dubke1988b} presented a theoretical model for describing
the electromagnetic forces and fluid flow occurring in electromagnetic stirring
of continuously cast strands with a rectangular cross-section. Experiments 
were carried out in a cold model for verifying the calculations. However, the 
flow velocity measurements performed rely on photographical techniques or a drag probe, 
respectively. Both approaches do not allow for quantitative flow measurements at high
spatial and temporal resolution.     

A numerical study of an RMF-driven flow in a square container was published 
by Frana and Stiller \cite{Frana2008}. The authors found that the velocity 
field is influenced by the corner effects and exhibits a non-axisymmetric 
structure in a wide range of magnetic Taylor numbers. In this study
the investigations are focused on the spin-up process resulting
from the application of the electromagnetic driving force in form of a step 
function to the fluid being at rest in the initial state. The magnetic field
was initiated at the time $t=0$ and induces a mainly azimuthal body force 
inside the liquid driving a primary swirling flow. The centrifugal force changes 
across the boundary layers at the horizontal walls of the container. This is
balanced by a radial pressure gradient resulting in a liquid flow towards 
the cylinder axis inside the horizontal boundary layers. This mechanism also
known as Ekman pumping is responsible for the existence of the secondary 
flow appearing as double vortex in the radial-vertical plane.

In this paper we present a combined experimental and numerical study devoted
to the transitional behaviour of the flow in a cubic container driven by a
rotating magnetic field. The paper is organized as follows:
the experimental setup and its instrumentation is presented in section
\ref{setup}. The governing equation for describing the induced 
electromagnetic field in the liquid melt and specially the derivation
of a semi-analytical expression for the time average of the induced
electromagnetic force density in the electrically conducting medium 
are described in detail in section \ref{section:numerics}.
In section  \ref{results} we present and discuss the main results of the numerical
simulation and the measurements in the experiments.
In order to obtain a better understanding of the transitions between 
different flow regimes we perform a Proper Orthogonal Decomposition (POD)
of both, the numerical simulation and the measurements.
This study leads to the identification of the characteristic exponential
growth of the first emerging flow instabilities.
The characteristic frequencies and growth rates are estimated.
The results are summarized in section \ref{conclusions}.

\section{Experimental setup}
\label{setup}
The experiments are conducted using the eutectic alloy GaInSn, enclosed in a 
cubic container made of acrylic glass with an edge length of 
$2 L = \unit{67.5}{mm}$.
Due to the low melting point, $\vartheta_m=\unit{10}{^\circ C}$, the metal is 
liquid at room temperature. The container is positioned in the center of the 
magnetic system MULTIMAG (MULTI purpose MAGnetic field) facility, which is capable 
of generating different types of magnetic fields of varying strength and 
frequency with high accuracy~\cite{Pal2009}. The magnetic system consists of
a radial arrangement of six coils, whereby opposing coils are connected
as pole-pairs.
A three-phase current generates a horizontal RMF rotating in the horizontal plane 
in clockwise direction with a frequency of $\unit{50}{Hz}$. The symmetry axes of 
magnetic field and fluid container are identical. 
The origin of the coordinates is collocated in the center of the cube.
Special care was taken to ensure 
a precise positioning of the cube inside the magnetic field for avoiding flow
artifacts caused by a misplacement of the fluid volume with respect to the
 magnetic field. 
The homogeneity of the magnetic field was checked using a 3-axis Gauss meter 
(Lakeshore model 560, sensor type MMZ2560-UH) and was found to be better than
3 \%.

The fluid velocity was measured using the ultrasound Doppler velocimetry (UDV),
which allows for determining instantaneous velocity profiles in opaque 
fluids \cite{Takeda1991}. The measuring principle is based on the pulse - 
echo technique and uses short ultrasound
bursts emitted from an ultrasound transducer,
that reflecte from acoustic inhomogeneities moving inside the fluid and are
 received from the transducer.
In case of GaInSn, these inhomogeneities are assumed to be microscopic oxide particles 
The measurements rely on the
assumption, that the reflecting particles are homogeneously dispersed in the melt 
and follow the flow without slip. Velocity profiles are reconstructed using information
about (i) the distance between the UDV-transducer and the reflecting particle which 
is derived from the time of flight of the ultrasound burst and (ii) the velocity
component along the beam direction which is determined from the phase shift of 
subsequent ultrasound burst echos.

\begin{figure}[htb]
  \includegraphics[width=0.62\textwidth]{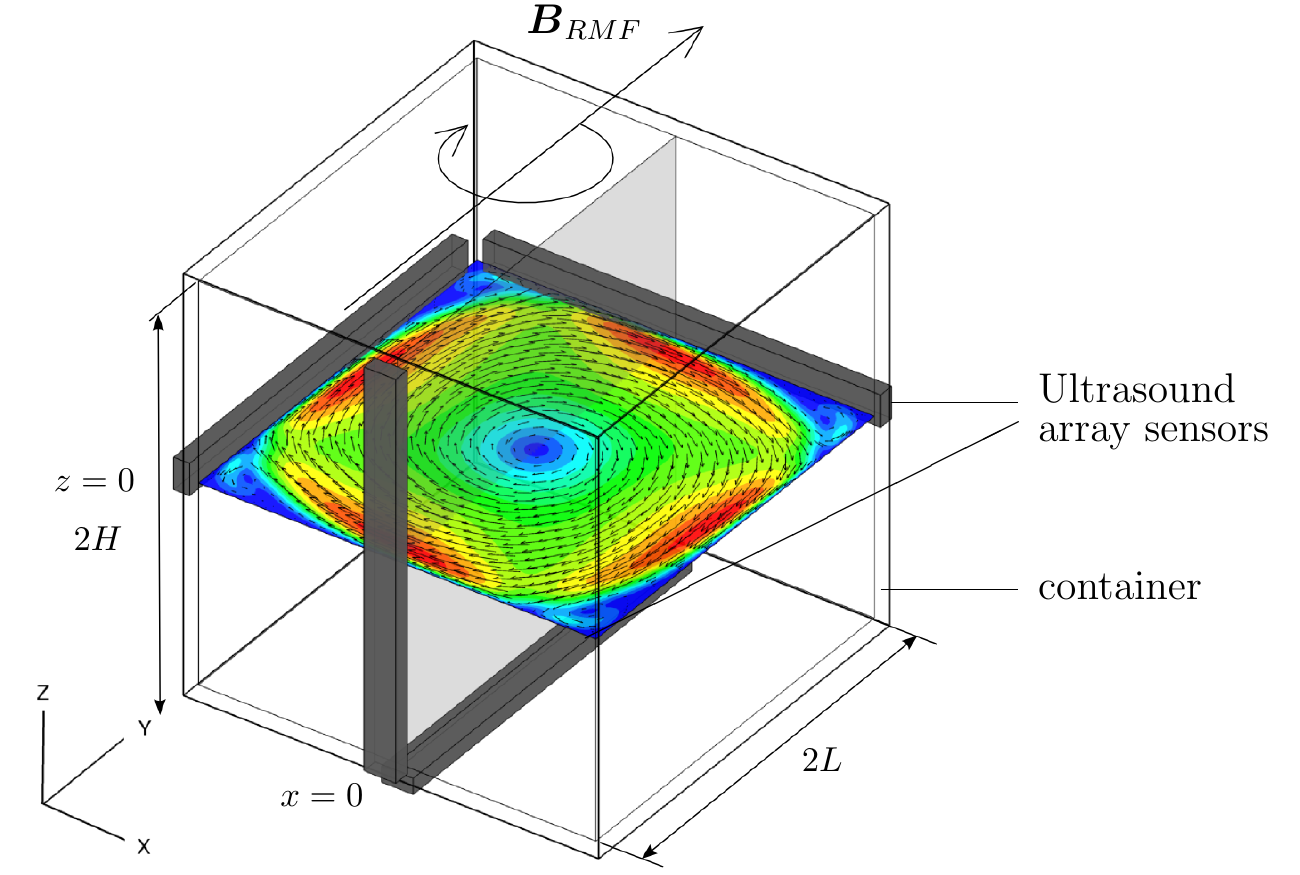}
  \vspace*{-0.5em}
\caption{Measurement configuration: a cubical vessel filled with GaInSn,
instrumented with four ultrasound array sensors. An example for a typical reconstructed vortex flow 
in the horizontal plane is shown here. The color represents the intensity of the flow and the arrows the 
direction in the plane.}
\label{fig:measconfig}
\end{figure}

Within this study we applied an ultrasound array Doppler velocimeter
(UADV) ~\cite{Franke2010, Nauber2013} 
in order to perform a two-dimensional flow mapping in both the horizontal and 
vertical mid-plane of the cube.
In our study the cube is instrumented
with four linear ultrasound arrays arranged orthogonally. Such an arrangement
provides a two-dimensional mapping of the two in-plane velocity components
(cf. Fig. \ref{fig:measconfig}).

\begin{figure}[htb]
  \includegraphics[width=0.57\textwidth]{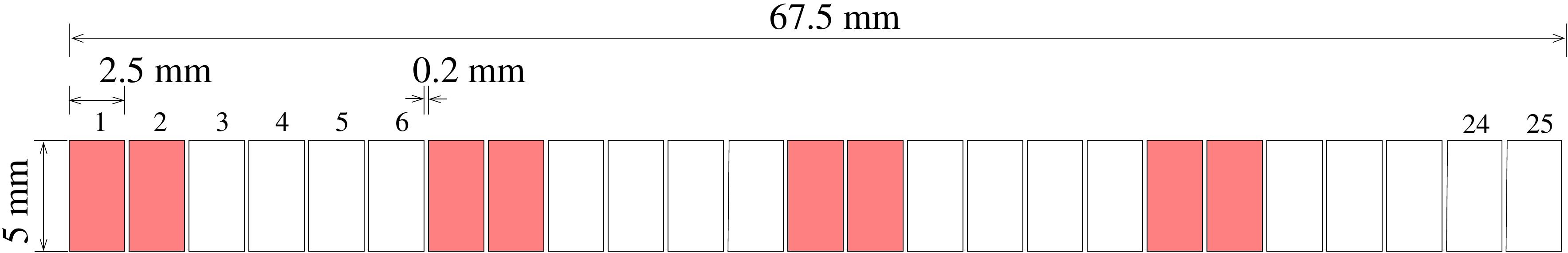}
\caption{Geometrical dimensions of the ultrasound array sensor,
the marked elements are active during the first scanning step.}
\label{fig:array}
\end{figure}

Each array consists of $25$ transducers with the dimensions of 
$\unit{2.5 \times 5}{mm^2}$ 
resulting in a total sensitive length of $\unit{67.5}{mm}$ (cf. Fig.~\ref{fig:array}). 
A pairwise excitation of neighboring elements gives an active surface of 
$\unit{5 \times 5}{mm^2}$ associated  with a sound beam width of approximately 
$\unit{3}{mm}$ in GaInSn.
The excitation signal is eight periods of a sine wave at $f=\unit{8}{MHz}$ 
resulting in an axial resolution of about $\unit{1.4}{mm}$~\cite{Franke2013}. 
The acoustical impedance of the transducers is matched to PMMA  
($\unit{3.4}{MRayl}$), which maximizes the sound transmission through the
sensor - wall interface.

In order to acquire a planar velocity map,
an electronic scanning of the respective linear sensor arrays is performed.  
The frame-rate is increased over a simple sequential scan by
parallelizing the measurement based on a combined space division/time division
multiplexing scheme~\cite{Franke2013}. 
In this way a frame-rate of up to $\unit{33}{Hz}$ can be achieved
in the given configuration.
To avoid crosstalk between the sensor arrays in this configuration, all four 
arrays are driven sequentially.

The velocity information is extracted from the amplified and digitized
ultrasound (US) echo 
signals via the Kasai autocorrelation method \cite{Kasai1985}. 
A typical mean data bandwidth after 
digitalization is $\unit{1.2}{GB/s}$, which is beyond the limit that can be 
acquired and stored continuously with common PC hardware. Therefore a real-time
data compression is performed by offloading parts of the signal processing to 
a field-programmable gate array (FPGA, NI PXIe-7965R). The preprocessing 
reduces the amount of data by 10:1 and enables a continuous streaming for a 
practically unlimited duration \cite{Nauber2016}.

\section{Governing equations}
\label{section:numerics}

Let us consider the flow of an electrically conducting fluid with kinematic 
viscosity $\nu$, density $\rho$ and electrical conductivity $\sigma$ in a cuboid 
container with the basis edge length $2 L$ and the height $2 H$ driven by a uniform 
magnetic field of induction $B_0$ rotating around the vertical axis $\bi e_z$ 
with a constant angular frequency $\omega$.

\subsection{Induced electromagnetic force}

In the scope of the low-induction approximation (very small magnetic Reynolds 
number $R_m=\mu_0 \sigma u_0 L << 1$) the electro-motive field $\bi u \times \bi B$ 
can be neglected compared to the induced electric field $\bi E$ within the Ohm's law 
$\bi j = \sigma ( \bi E + \bi u \times \bi B)$. Here is $\mu_0$ the magnetic 
vacuum permeability and $u_0$ is a characteristic velocity of the flow. The back 
reaction of the flow field $\bi u$ on the total induced electric current 
density $\bi j$ can be neglected, too. This is why the simulations of the 
electromagnetic field and the fluid flow can be conducted separately.
Hence, a quasi - analytical expression for the time-averaged electromagnetic force 
density $\bi j \times \bi B$ acting on the liquid metal in the cavity can be derived. 

A clockwise rotating magnetic field with strength $B_0$ can be expressed as:

\begin{equation} \label{eqnb}
\bi B = B_0 \left\{ cos ( \omega t) \bi e_x - sin ( \omega t) \bi e_y
              \right\} \, .
\end{equation}

The corresponding magnetic vector potential 
($\bi B = \nabla \times \bi A$) has only one axial component: 
$\bi A = B_0 \left( y \, cos ( \omega t) + x \, sin ( \omega t)
              \right) ~ \bi e_z
$. The unit vectors $\bi e_x$, $\bi e_y$ and $\bi e_z$ are
related to the axes of the reference system and their orientation is
sketched in Fig. \ref{fig:measconfig}.

Within the considered approximation the electric current density
can be calculated using the Ohm's law and the first Maxwell's equation:

\begin{equation} \label{eqnj}
\bi j = \sigma \bi E = \sigma \{ - \nabla \Phi -
          \frac{\partial \bi A}{\partial t} \}
        = \sigma \left\{ - \nabla \Phi -
          B_0 \omega \left( - y \, sin ( \omega t) + x \, cos ( \omega t)
                 \right) \bi e_z
                 \right\} \, .
\end{equation}

Here is $\Phi$ the electric potential.
The induced currents have been neglected in the frame of the low frequency
approximation assuming a complete penetration of the fluid volume by the
 magnetic field.
In a next step we compute the instantaneous electromagnetic force density 
taking into account the expressions \ref{eqnb} and \ref{eqnj}:

\begin{eqnarray} \label{jxb}
\bi j \times \bi B & = &
\sigma B_0 \{
( - \frac{ \partial \Phi}{\partial z} - \omega B_0
           (- y sin(\omega t)+ x cos(\omega t) ) )
           ( sin(\omega t) ~ \bi e_x + cos(\omega t) ~ \bi e_y )
           \nonumber
\\ & &
           +( \frac{\partial \Phi}{\partial y} cos(\omega t)
             + \frac{\partial \Phi}{\partial x} sin(\omega t)
            )  \bi e_z
           \} \, .
\end{eqnarray}

Until this point the equation \ref{jxb} shows the same form as presented in the paper
from Frana and Stiller \cite{Frana2008}. 
Now we obtain a semi-analytical expression for the electromagnetic force density
using a proper ansatz for the electric potential $\Phi$:

\begin{equation} \label{ansatz}
  \Phi(x,y,z,t) = \omega B_0 ( a(x,y,z) cos(\omega t) + b(x,y,z) sin(\omega t)
                             ) \, .
\end{equation}

The functions $a(x,y,z)$ and $b(x,y,z)$ have the dimensions $\mathrm{m}^2$. Whilst taking
into account that $ 1/T \int_0^T dt~sin^2(\omega t) = 1/T \int_0^T dt~ cos^2(\omega t) = 1/2$ we perform now the time averaging of the force density
defined in Eq. \ref{jxb} 
over a period $T=2 \pi/\omega$ in order to separate its steady part:

\begin{equation}
\bi f = 
< \bi j \times \bi B >_T = \frac{\sigma \omega B_0^2}{2}
 \left\{ ( - \frac{\partial b}{\partial z} + y )~ \bi e_x
       + ( - \frac{\partial a}{\partial z} - x )~ \bi e_y
       + ( \frac{\partial b}{\partial x}
         + \frac{\partial a}{\partial y}  )~ \bi e_z
 \right\} \, .
\end{equation}

The Kirchhoff's law for the electric charge conservation $ \nabla \cdot \bi j = 0$ 
leads to the equation for the electric potential $\nabla^2 \Phi = 0$.
Using the ansatz \ref{ansatz}, we obtain the following system of equations:

\begin{equation} \label{eqnab}
\nabla^2 a = 0 \quad \mbox{and} \quad \nabla^2 b = 0
\end{equation}

which should be solved under the boundary condition excluding any electrical 
current flowing through isolating walls ($j_n = 0$).
It means, for example, for the functions $a$:

\begin{equation} \label{bcab}
\frac{\partial a}{\partial z} |_{z = \pm H} = - x \quad \mbox{and} \quad
\frac{\partial a}{\partial x}|_{x = \pm L}  =
\frac{\partial a}{\partial y}|_{y = \pm L}  = 0 \, .
\end{equation}

Due to reasons of symmetry, it can be shown that $a(x,y,z) = a(x,z)= - b(y,z)$. 
This fact leads to the following final expression for the time averaged
electromagnetic force density:

\begin{equation} \label{femd}
\bi f = \frac{\sigma \omega B_0^2}{2}
 \left\{ (\frac{\partial a}{\partial z}(y,x,z) + y )~ \bi e_x
       + ( - \frac{\partial a}{\partial z}(x,y,z) - x )~ \bi e_y
 \right\} \, .
\end{equation}

The functions $a$ and $b$ are solutions of the equation \ref{eqnab} on the 
boundary conditions \ref{bcab}.
Using $L$ as length scale and $\rho \nu^2/L^3$ as the scale for the force 
volume density we can express equation \ref{femd} in a dimensionless form 
$\bi f = Ta~ \bi f_{EM}$ with

\begin{equation} \label{fem}
  \bi f_{EM} = r~ \bi e_\varphi +
  \left( - \frac{\partial b}{\partial z} ~ \bi e_x
    - \frac{\partial a}{\partial z} ~ \bi e_y
  \right) \quad \mbox{and} \quad
  Ta=\frac{\D \sigma  \omega  B_0^2 L^4}{\D 2 \rho \nu^2} \, .
\end{equation}

$Ta$ denotes the magnetic Taylor number, which is an expression for the
relative strength of the electromagnetic force driving the flow.

The first term in equation \ref{fem} is strictly azimuthal and linear in the 
polar radius $r=\sqrt{x^2+y^2}$ and the second term takes into consideration 
the geometry of the cross section and the finite height of the container. 

\begin{figure}[htb]
\begin{center}
  \includegraphics[width=0.825\textwidth]{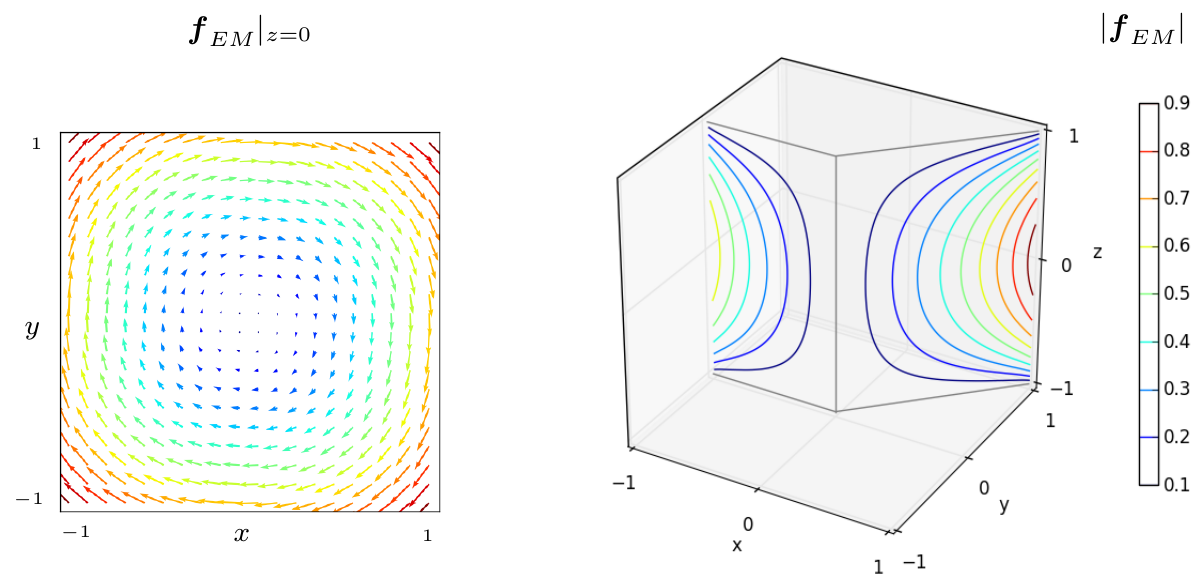} 
\end{center}
  \vspace*{-1em}
\caption{\label{lorentz}Simulation: Vector plot (left) and 
amplitude (right) of the non-dimensional electromagnetic force 
density $\bi f_{EM}$ (Eq. \ref{fem}) at different sectional planes.}
\end{figure}

Fig. \ref{lorentz} shows the spatial distribution of the time-averaged non-dimensional
electromagnetic force density $\bi f_{EM}$, which was added as an additional body 
force to the Navier Stokes equation for the case of an aspect ratio $H/L = 1$.
The force reaches its maximum value $\bi f_{{EM}_{max}}=0.953$
at half height in a vertical edge of the cube, i.e. for example at the position
with the coordinates $x=1,y=1,z=0$.

The electromagnetic force density distribution for arbitrary aspect ratios
can be determined by the procedure described herein.
A study of the RMF induced flow for different aspect ratios will be published 
in the future, however, this paper focuses on the aspect ratio $H/L = 1$,
which is realized in the experimental setup.

\subsection{Electromagnetically driven flow}

The numerical simulations of the liquid metal flow are performed using the
open source code 
library OpenFOAM$^\copyright$ 3.0.x \cite{openfoam}.
The flow was computed solving the incompressibility condition $\nabla \cdot \bi u = 0$ 
and the incompressible Navier-Stokes equation in 
a dimensionless form, with $L$, $L^2/\nu$ and $\rho (\nu/L)^2$ being the distance,
time and pressure scale, respectively, is given by

\begin{equation} \label{ns}
\frac{\partial \bi u}{\partial t} + (\bi u \cdot \nabla ) \bi u
= - \nabla p + \nabla^2 \bi u + Ta~ \bi f_{EM} \, .
\end{equation}

The no-slip condition $u=0$ at the solid container walls was chosen as boundary 
condition for the calculation of the flow field.

The set of equations is solved using the PISO (Pressure Implicit with
Splitting Operator) algorithm \cite{issa1986,ferziger2003} on a collocated grid.  The time step was chosen so
that the Courant number always remains below 0.25. We use a structured
numerical grid with two million volume elements and refinements near the walls.
Second order accurate schemes are used for time (Crank-Nicolson) and space
discretization (linear scheme).

In order to obtain mesh refinement independent solutions reliably, we performed
mesh convergence studies using 4 different discrete meshes with $10^6$ (mesh I),
$1.57 \times 10^6$ (mesh II), $2 \times 10^6$ (mesh III) and $2.46 \times 10^6$
(mesh IV) number of cells, respectively. For the mesh generation we used the
standard blockMesh OpenFOAM tool with multi-grading. We stretched the mesh 
between $x=0.9$ and $x=1$ (wall) with a stretching factor of 
$\Delta x_{max}/\Delta x_{min}=5$. Doing so, the maximum volume aspect ratio 
remain less than 0.5. The smallest cell edge length is $\Delta x_{min}=0.00329$
and the boundary layer region between $x=0.9$ and $x=1$ contains 12 cells.
In Sec. \ref{transitiontounsteadyflow}, we show the results of the convergence studies at different
typical flow regimes.

\section{Results and discussion}
\label{results}

In this section we present numerical and experimental results with respect to the 
flow of GaInSn driven by a rotating magnetic field with strength $B_0$ in a 
closed cube with edge length $2 L=\unit{67.5}{mm}$
For the numerical simulations the following material properties 
of Ga$_{67}$In$_{20.5}$Sn$_{12.5}$ in \mbox{wt. \%} at $\unit{20}{^\circ C}$  
were used\cite{Plevachuk2014}:

\begin{center}
\begin{tabular}{l|lll}
density   & $~\rho$ : & $6403$ & $\unit{}{kg/m^3}$ \\
viscosity & $~\nu$  : & $0.34 \times 10^{-6}$ & $\unit{}{m^2/s}$ \\
conductivity & $~\sigma$ : & $3.29 \times 10^{6}$ & $\unit{}{S/m}$ \\
\end{tabular}
\end{center}

This results in a viscous time scale of $t_0=L^2/\nu=\unit{3341.23}{s}$ and a
velocity scale $u_0=\nu/L=\unit{0.0101}{mm \, s^{-1}}$.
Everything that is presented in this paper from now on will be given in a
dimensionless form taking into account the previously introduced scaling.

\subsection{Steady flow regime}

Below the critical value of the magnetic Taylor number $Ta_{c}$ the 
electromagnetically driven flow remains laminar and steady.
In Sec. \ref{transitiontounsteadyflow},
we will identify this critical value by analyzing both the time-dependent 
numerical simulations and the flow measurements carried out during 
the experiments. In both cases, the flow becomes 
unsteady beyond $Ta > 1.3 \times 10^5$.

For a better understanding and a better description of the development of
the main flow structures we split the velocity field $\bi u$ in two parts:
the so-called primary flow,
which contains the azimuthal component only $u_{\varphi} \bi e_\varphi$
and the remaining part $\bi u - u_{\varphi} \bi e_\varphi$, which we call
secondary flow.

\begin{figure}[hbt]
\begin{center}
  \includegraphics[width=0.9\textwidth]{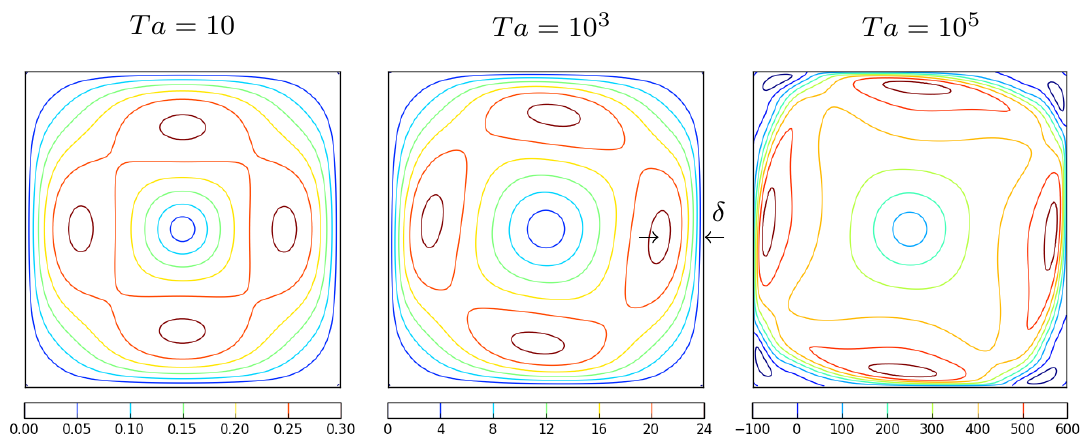}
  \vspace*{-1.1em}
\end{center}
\caption{\label{vxy} Simulation: Contours of the azimuthal flow velocity
$- u_\varphi$ at the middle horizontal plane $z=0$
show the decreasing of the boundary layer size
$\delta$
and the loss of the mirror symmetry with respect to the
$x$ and $y$ planes for increasing values of the Taylor number $Ta$.
        }
\end{figure}

In general, the flow structure changes rapidly for increasing driving force strength. 
For very small values of the Taylor number, i.e. $Ta < 10$, the flow structure 
shows mirror symmetries with respect to the $x$,$y$ and $z$ planes (cf. Fig. \ref{vxy}).

The increase of the Taylor number causes a remarkable deformation of
the contour lines of the azimuthal flow velocity 
$u_\varphi = − u_x sin(\varphi) + u_y cos(\varphi)$.
The higher the chosen $Ta$, the narrower the boundary
layers of the azimuthal velocity are formed in the vicinity of the
container walls. This represents a challenge for both the numerical
simulation and the measurements with respect to ensuring an appropriate
resolution near the wall. Within this study the use of stretched meshes
in the numerical simulations guarantees a sufficient resolution of
the boundary layers.
Starting from the center axis the azimuthal
velocity increases linearly with the distance from the vertical axis.
It achieves a maximum value at certain position $\bi r_m$ and
finally it decreases to zero value at the side wall. 
The thickness of the boundary layer for the azimuthal velocity
can be defined as the
distance between the place where the azimuthal velocity has a maximum
and the side wall, i.e. $\delta = 1- \bi r_m \cdot \bi e_x$.

Fig. \ref{vxy} reveals that for high values of the
magnetic Taylor number $Ta$ (i.e. for $Ta>6 \times 10^3$) small regions
in the vicinity of the  corners appear where the fluid rotates in a
direction opposite to the main rotation direction. This causes the
shear to increase and thus the tendency for the formation of flow
instabilities.

\begin{figure}[hbt]
\begin{center}
  \hspace*{-1em}
  \includegraphics[width=0.95\textwidth]{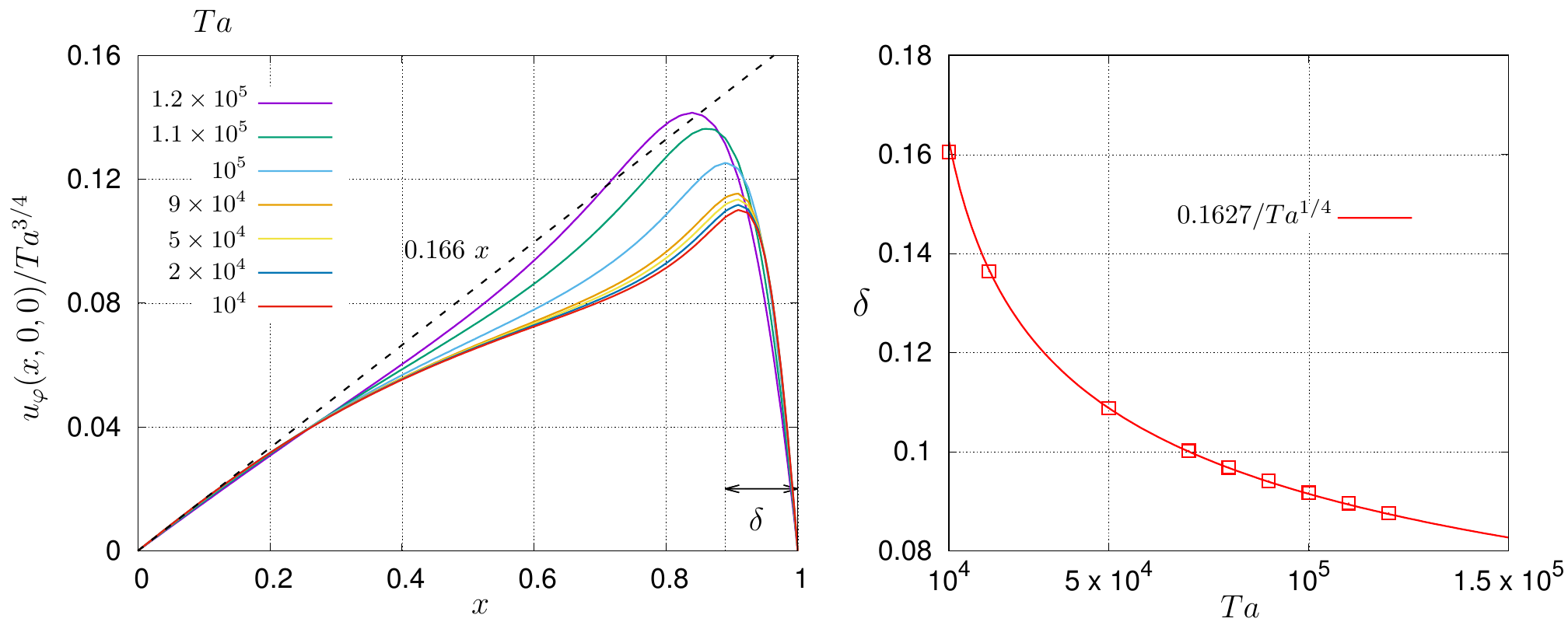}
\end{center}
\vspace*{-1.1em}
\caption{\label{uphivsx} Simulation: Azimuthal velocity profiles
at the middle horizontal plane (left) and boundary layer thickness
$\delta$ (right) for different values of the Taylor number
}
\end{figure}

Fig. \ref{uphivsx} shows the azimuthal velocity profiles at the middle horizontal
plane ($u_\varphi(x,0,0)$, left) and the boundary layer thickness
$\delta$ (right) for different values of the Taylor number.
The latter scales like $\delta = 1.627~ Ta^{-1/4}$.
Applying the scale $Ta^{3/4}$ for the
azimuthal velocity, the core rotation speed $\Omega=u_\varphi/r|_{r \to 0}$
shows a self similar behaviour, i.e.
$\Omega = 0.166 ~ Ta^{3/4}$ is obtained for $0.2\times 10^4 < Ta < 1.5 \times 10^5$
in a contrast to different scaling behaviours discussed in \cite{Niki2005}
for the case of the RMF flow in a cylindrical container
($\Omega \propto Ta^{2/3}$ in that case).
This means for example in physical units, that for $Ta=10^5$ the central
core rotates with an almost constant angular velocity
$\Omega = \unit{0.278}{s^{-1}}$ while the RMF does this with the
angular frequency $\omega = 2 \pi \times  \unit{50}{s^{-1}}$.
The boundary layer thickness for $Ta=10^5$ is 
$\delta = 0.0917$. Within this region the grid is stretched,
it contains 11 cells and the smallest cell edge length is 
0.00358 in non-dimensional units. This fact guarantees a correct description
of the boundary layer.

\begin{figure}[htb]
\begin{center}
  \includegraphics[width=0.61\textwidth]{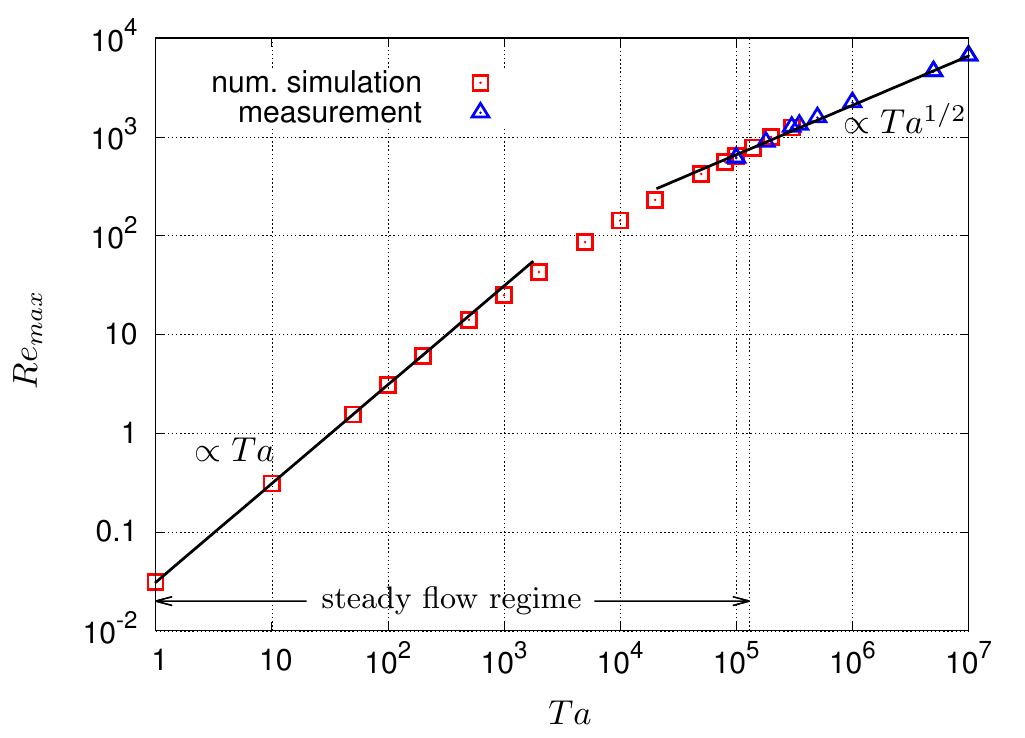}
  \vspace*{-1.1em}
\end{center}
\caption{\label{umax} Scaling behaviour: Maximum Reynolds number of the primary
flow $Re_{max}$ (cf. Eq. \ref{remax})
as a function of the magnetic Taylor number $Ta$.
        }
\end{figure}

Let us examine the scaling behaviour of the maximum velocity
of the primary flow at the central horizontal plane $z=0$
with respect to the magnetic Taylor number $Ta$.
For an infinitely long circular cylinder an analytical expression for
the velocity field exists under the assumptions that the flow is laminar
and that the velocity field has only one azimuthal component depending
on the polar radius only: $ u_\varphi(r) = r ~ (1-r^2)~ Ta/8$.
The maximum velocity value
${u_\varphi}_{max} = Ta/(12 \sqrt{3})$ is directly proportional to the
Taylor number $Ta$. 
In the case of the RMF-driven flow in a cubic container as studied here,
the azimuthal velocity profile along a horizontal line exhibits
a similar behaviour.
Figure \ref{umax} shows the $Ta$ number dependence of the maximum 
Reynolds number associated with the primary flow 

\begin{equation} \label{remax}
  Re_{max} = max \{ \sqrt{(u_x^2+u_y^2)}|_{z=0} \}
\end{equation}

from the numerical simulation and the measurements.

Figure \ref{umax} reveals two characteristic asymptotic
scaling laws.  A linear relationship
appears for small values of the Taylor number ($Ta < 2 \times 10^2$): 
$Re_{max} \approx 0.031~ Ta$. The corresponding relation for the laminar flow 
in an infinitely long circular cylinder is rather similar, namely
$Re_{max} = Ta/(12 \sqrt{3})$. 
For $Ta > 4 \times 10^5$ we find 
the relationship $Re_{max} \approx 2.08~ Ta^{1/2}$ in concordance with the
predictions made by Davidson and Hunt \cite{davidsonhunt1987}.

\begin{figure}[htb]
\centering
  \includegraphics[width=0.75\textwidth]{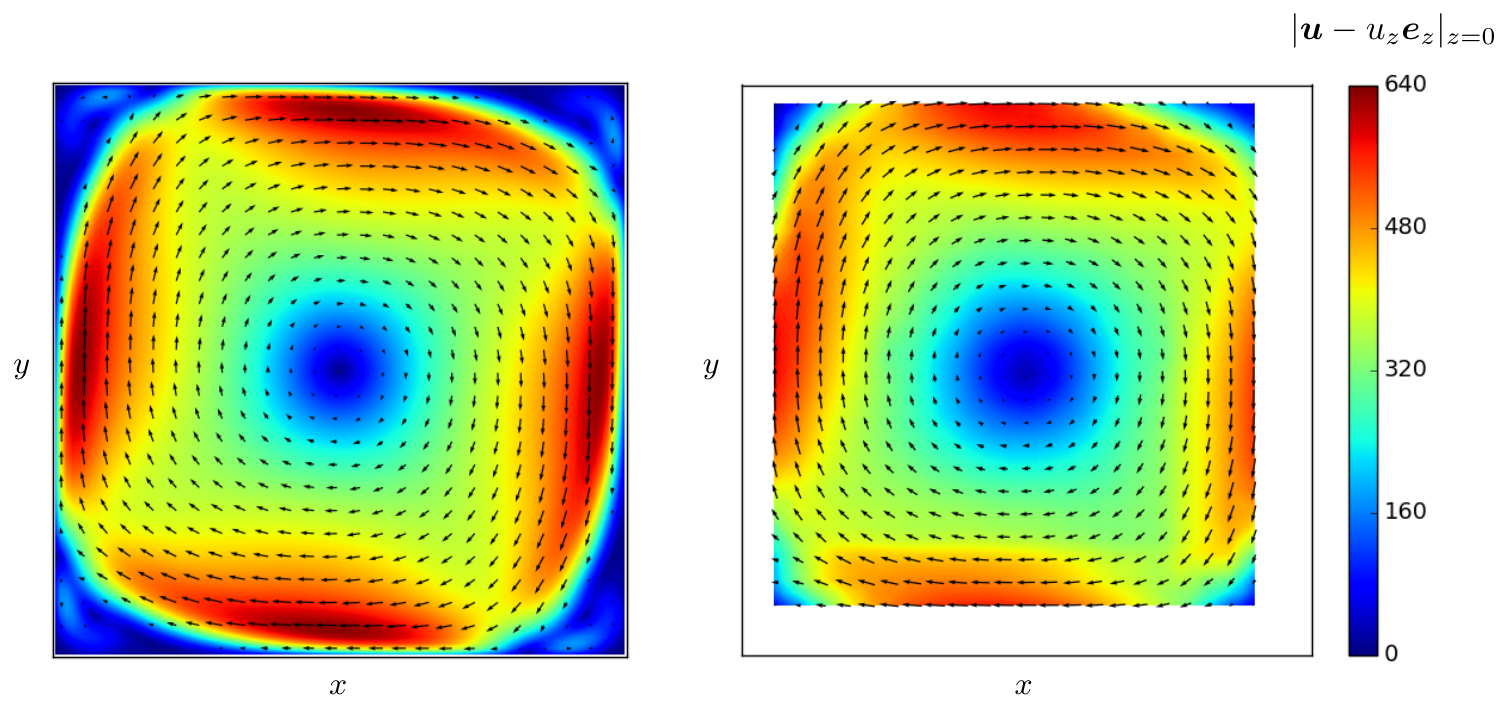}
  \vspace*{-0.6em}
\caption{\label{primaryflow}
Velocity distribution of the primary flow ($u_x \bi e_x + u_y \bi e_y)|_{z=0}$
for $Ta=10^5$ (left: computed steady state from the numerical
simulation, right: measured mean values)}
\end{figure}

Fig. \ref{primaryflow} shows velocity distributions of the primary flow 
in the horizontal plane $z=0$ (i. e., $(u_x \bi e_x +u_y \bi e_y)|_{z=0}$)
computed from the numerical simulation (left) and measured during the experiments 
(right) for $Ta=10^5$. The measured values presented here were averaged over the
time interval $1000 - \unit{2000}{s}$.
This direct comparison reveals a very good agreement. The flow pattern reconstructed
from the UADV measurements by interpolation does not fill the entire cross section. 
Due to the dimensions of the sensor housing and constructive limitations of the 
container design the flow field cannot be acquired in the immediate vicinity of the
vessel walls, especially not directly at the walls where the arrays are installed 
(bottom and right side in the right part of Fig. \ref{primaryflow}). 
The measuring lines of the outmost transducers within the linear arrays run in a 
distance of \unit{2.7}{mm} parallel to the walls. Moreover, the reverberation 
of the transducers and the US transmission through the channel wall result 
in a saturation of the transducer preventing measurements at depths located just 
a few millimeters behind the inner wall.         

\begin{figure}[htb]
\centering
  \includegraphics[width=0.75\textwidth]{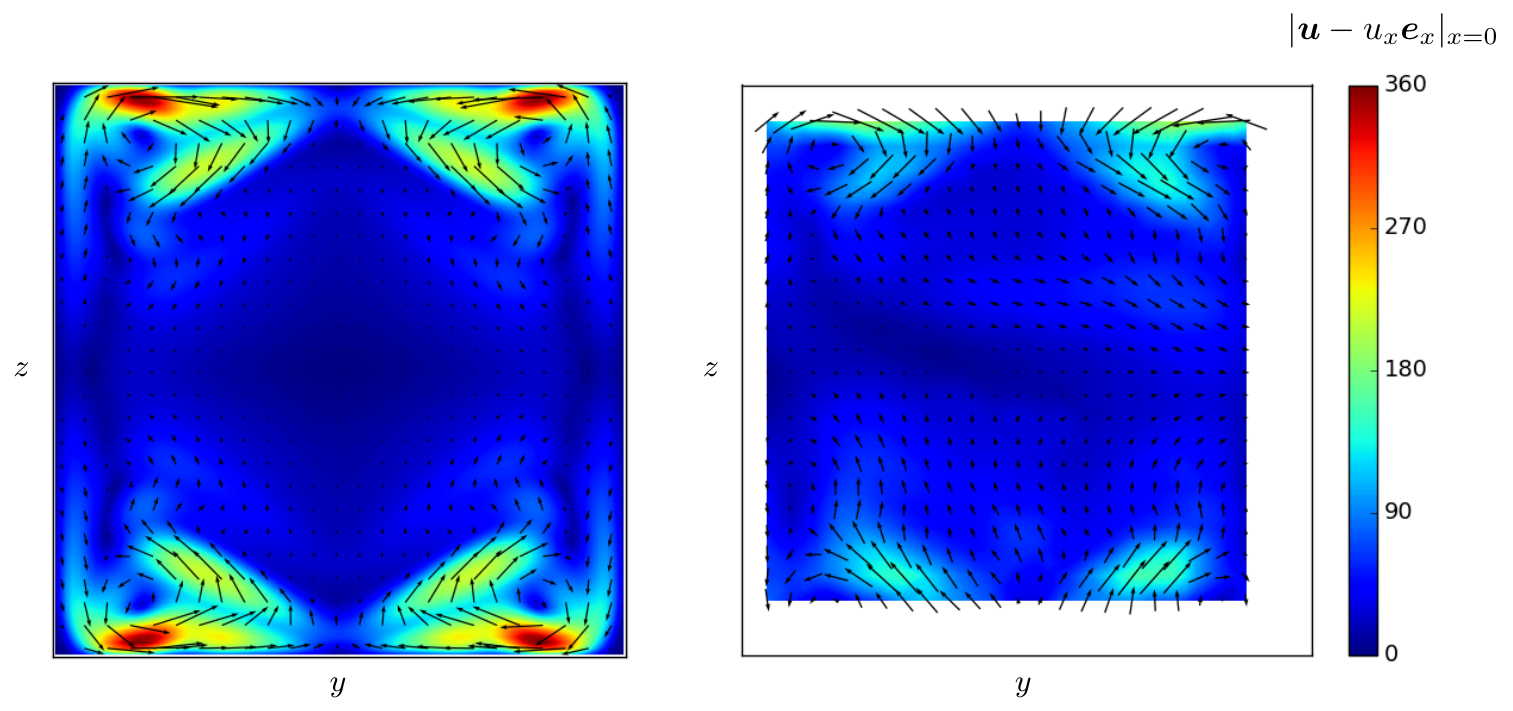}
  \vspace*{-0.6em}
\caption{\label{uyzcompare}
Velocity distribution of the secondary flow ($u_y \bi e_y + u_z \bi e_z)|_{x=0}$ 
for $Ta= 10^5$
(left: computed steady state from the numerical simulation, right:
measured mean values)
}
\end{figure}

A corresponding comparison of numerical simulation and experiment with respect to
the secondary flow can be found in Fig. \ref{uyzcompare} showing the flow pattern
in the meridional plane $x=0$ for $Ta=10^5$. Due to the saturation effects 
described above, the measurement system does not provide valid velocity measurements 
near the walls. Unfortunately, the secondary flow shows the most striking structures 
and maximum values just in regions near the top and the bottom of the container,
which are out of the UADV measuring domain. The numerical results reveal the
existence of distinct vortex pairs which are driven by Ekman pumping in the 
top and bottom boundary layers.

\begin{figure}[htb]
\centering
  \includegraphics[width=0.6\textwidth]{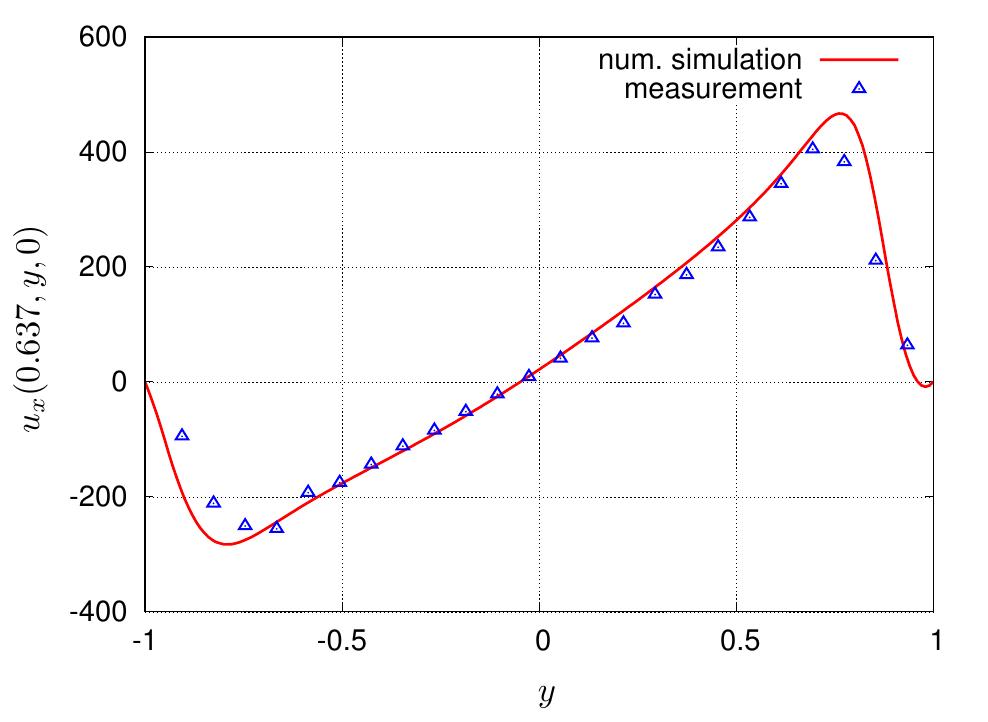}
  \vspace*{-1em}
\caption{\label{profile} Mean measured profile of the
velocity component $u_x$ (blue points) and 
corresponding numerical simulation profile (red line)
along the line $x=0.637, -1<y<1, z=0$ for $Ta=10^5$.}
\end{figure}

In order to validate the numerical scheme we compared the numerical results
and the measured experimental data with respect to both the velocity profiles
and the integral quantities as the kinetic energy of the secondary flow.
Figure \ref{profile} presents an example of a mean profile
of the velocity component $u_x$ along the line $x=0.637, -1<y<1,z=0$
for $Ta=10^5$.

\subsection{Transition to unsteady flow regimes}
\label{transitiontounsteadyflow}

In order to assess the critical Taylor number $Ta_{c}$ for the transition from
laminar to oscillatory or unsteady flow regimes, we consider the time evolution
of the velocity field during the spin-up process where the flow field is evolving
from the state of rest after a sudden start-up of the RMF.

\begin{figure}[htb]
\centering
  \includegraphics[width=0.6\textwidth]{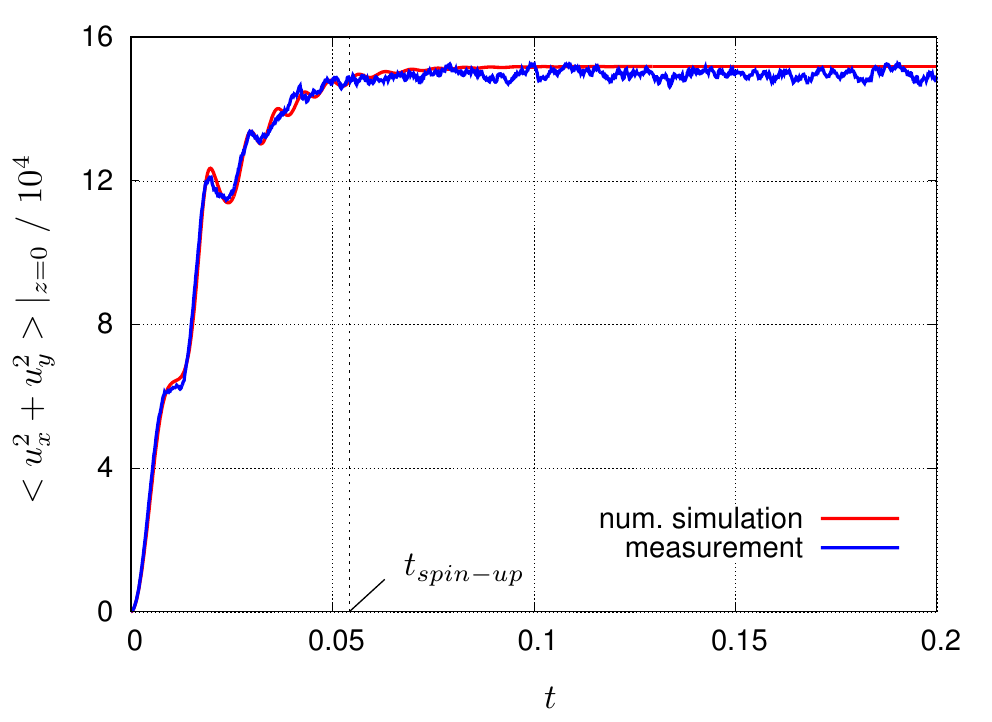}
  \vspace*{-1em}
\caption{\label{spinup}Spin-up of the kinetic energy of the primary
flow $<u_x^2+u_y^2>$
at the middle horizontal plane $z=0$ for $Ta=10^5$ (blue: measured 
and red: values computed from the numerical simulation). 
$t_{spin-up}$ denotes the so-called spin-up time as defined
by Nikrityuk et al.  \cite{Niki2005} for RMF driven flows in a circular
cylindrical container.
}
\end{figure}

Fig. \ref{spinup} displays the time evolution of the kinetic energy of the
primary flow

\begin{equation}
   <u_x^2+u_y^2>|_{z=0} = \frac{1}{N_x N_y} \sum_{i,j}^{N_x,N_y} {u_x}_{ij}^2+{u_y}_{ij}^2
\end{equation}

in the middle horizontal plane $z=0$ for $Ta=10^5$.

The so-called spin-up time
$t_{spin-up} = 2 H/\sqrt{\nu \Omega_{ce}} \propto Ta^{-1/3}$
is utilized to analyze spin-up dynamics of a melt driven by RMF in a circular
cylindrical container \cite{Ungarish1997,Niki2005}. $\Omega_{ce}$ denotes the
effective steady-state angular velocity at the center. Using the expression (16)
given by Nikrityuk \cite{Niki2005} and applying it for $Ta=10^5$ we obtain: 
$t_{spin-up}=0.0542$, which is indicated for comparison in Fig. \ref{spinup}.

\begin{figure}[htb]
\begin{center}
  \includegraphics[width=0.625\textwidth]{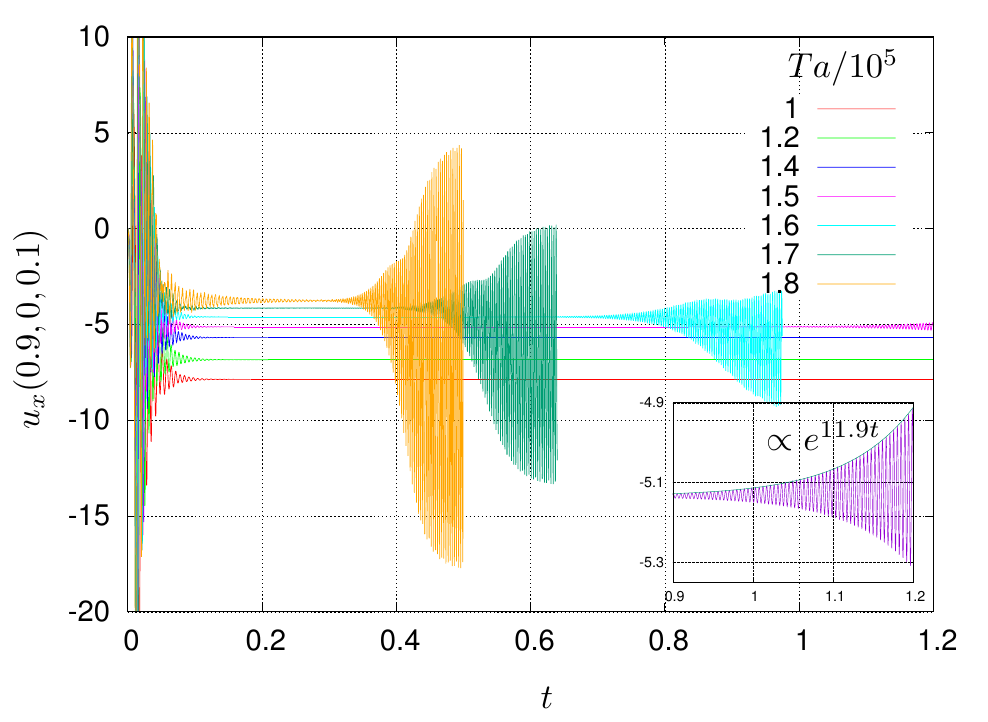}
  \vspace*{-1.5em}
\end{center}
\caption{\label{ux}Simulation: Time evolution of the velocity component $u_x$
         at the monitoring point $x=0.9, y=0, z=0.1$ for
         different values of the Taylor number $Ta$ showing the characteristic
         exponential growth of the first emerging flow instabilities.
        }
\end{figure}

Fig. \ref{ux} shows 
numerical results with respect to the time evolution of the velocity component $v_x$ 
at a monitoring point with the coordinates $x=0.9, y=0, z=0.1$ 
for different values of the magnetic Taylor number $Ta$. 

\begin{figure}[htb]
\centering
  \includegraphics[width=0.825\textwidth]{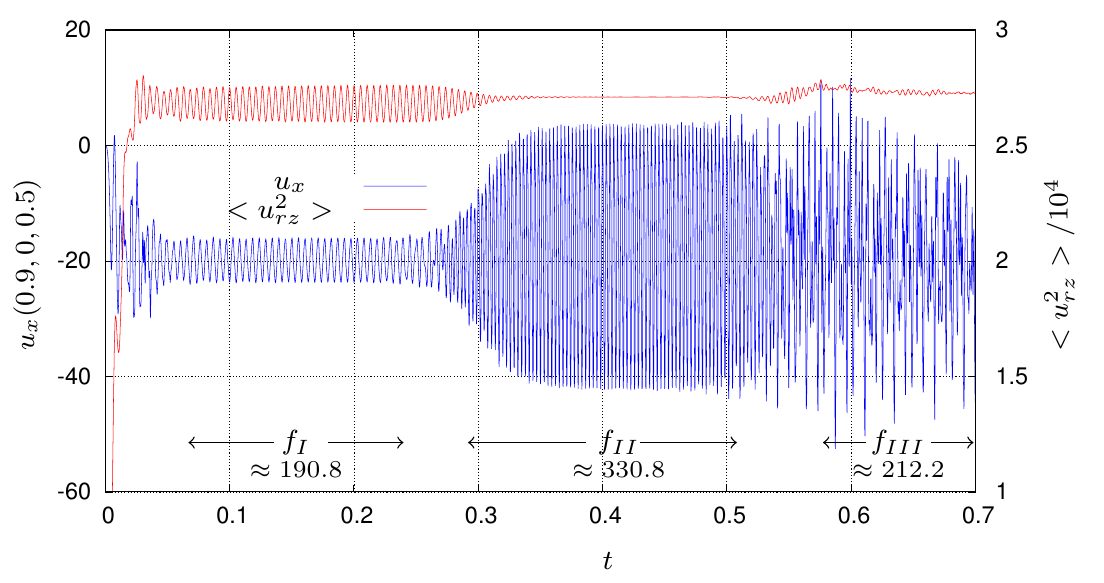}
  \vspace*{-1em}
\caption{\label{threefreq} Simulation: Time evolution of the kinetic energy of
the secondary flow (red) and of the velocity component $u_x$ at the 
monitoring point with the coordinates $(0.9, 0, 0.5)$ (blue) for
$Ta=1.9 \times 10^5$ showing 
three characteristic oscillation frequencies ($f_I$, $f_{II}$ and $f_{III}$)
occurring at three different time intervals, respectively.
 }
\end{figure}

The initial state is dominated by inertial oscillations which are forced by the 
rapid increase in the rotation rate. Because of viscous damping, these oscillations
decay and can be observed therefore only during a finite initial time period 
which roughly 
corresponds to the spin-up time ~\cite{Niki2005, Raebiger2010}. Finally, the flow 
reaches a steady state. The transition to a time-dependent flow regime becomes 
obvious by a reappearance of pronounced oscillations of the velocity signal. 
Figure \ref{ux} demonstrates the exponential growth of the instabilities. 
In this figure the drawings of the velocity curves for each Taylor number are terminated
at the time, where the saturation was achieved after the exponential growth phase.
We will discuss different transition phases later in this section 
in more detail (see Fig. \ref{threefreq}).
For $Ta = 1.6\times10^5$ we detected
using a proper orthogonal decomposition (see Sec. \ref{section:pod})
for the first upcoming instability
a growth rate of $\approx 20.5$ and afterwards a phase 
with velocity oscillations having the period of $ \approx 0.00345$.

The transition from a steady to a time-dependent flow regime was observed
for magnetic Taylor numbers being larger as a minimum value in the range of 
$1.2 \times 10^5< Ta_{c} < 1.3 \times 10^5$. The corresponding critical value for
the case of a finite circular cylinder of aspect ratio 1 is 
$Ta_{c} = 1.232 \times 10^5$ given by Grants et al. \cite{Grants2001}.

\begin{figure}[htb]
\begin{center}
  \includegraphics[width=0.675\textwidth]{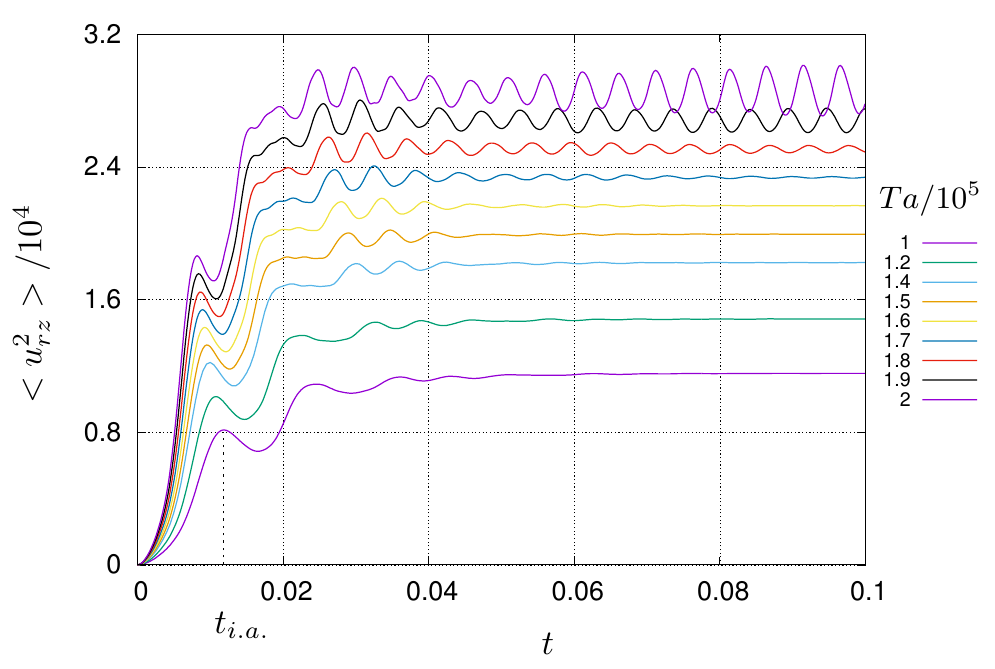}
  \vspace*{-1.75em}
\end{center}
\caption{\label{u2rz}Simulation: Time evolution of the mean kinetic energy
of the secondary flow (c.f. Eq. \ref{urz}). The time required for achieving
the first maximum is called initial adjustment time $t_{i.a}$.}
\end{figure}

Fig. \ref{u2rz} depicts the time evolution of the over the whole volume averaged
kinetic energy of the secondary flow

\begin{equation} \label{urz}
   <u_{rz}^2> = \frac{1}{V} \int dV \, ( u_r^2+u_z^2 ) \, ,
\end{equation}

where $u_r = u_x cos(\varphi) + u_y sin(\varphi)$ is the radial velocity 
component in polar coordinates.

The behaviour shown here appears to be similar as reported for the case of
 the RMF-driven flow in a finite cylinder 
\cite{Raebiger2010,Niki2005,Ungarish1997}. 

\begin{figure}[htb]
\centering
  \includegraphics[width=0.6\textwidth]{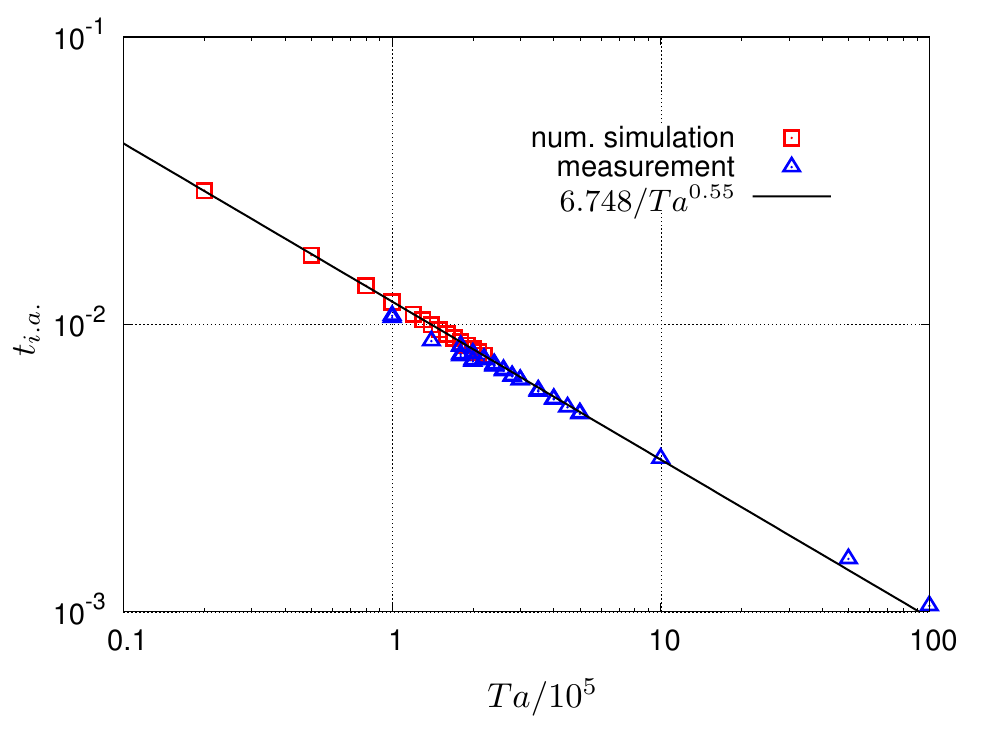}
  \vspace*{-1em}
\caption{\label{tia}Initial adjustment time $t_{i.a.}$ as a function of
the Taylor number $Ta$.}
\end{figure}

The time between initiating the magnetic field and reaching the first maximum 
of the energy amplitude is the so-called initial adjustment time $t_{i.a.}$ \cite{Niki2005}.
In Fig. \ref{tia} the initial adjustment time is drawn as a function of the
Taylor number $Ta$. The expression $t_{i.a.} = 6.748 /Ta^{0.54}$ describes
very good this relationship.

For Taylor numbers just above the critical value, i.e. for
$Ta < 1.4 \times 10^5$, the numerical simulations reveal a
convergence of the flow to a stationary state. A further increase
of the magnetic Taylor number leads to oscillatory flow regimes 
which characteristic frequencies depend on the magnetic Taylor number 
$Ta$ and the time passed since the initial adjustment time 
$t_{i.a.}$, respectively.
Fig. \ref{threefreq} helps to clarify these situations.

Fig. \ref{threefreq} presents both the time evolution of the
volume averaged kinetic 
energy of the secondary flow $<u^2_{rz}>$ (c.f. Eq. \ref{urz}) and the
velocity component $u_x$ at the monitoring point $P=(0.9,0,0.5)$
for $Ta=1.9 \times 10^5$. Unlike as in Fig. \ref{ux} here we also show the
flow behaviour beyond the onset of the saturation phase.
We can recognize three different oscillatory regimes at 
different moments in time showing unique characteristic
frequencies in each case, namely: $f_{I}$, $f_{II}$ and $f_{III}$ 
for the time intervals: $0.06 < t < 0.25$, $0.3 < t < 0.5$
and $ 0.58 < t$, respectively. The successive appearance
of different frequencies is closely related to the transient behaviour of
different flow modes. This can be studied in detail by a proper orthogonal
decomposition, which will be presented in section \ref{section:pod}.

\begin{figure}[htb]
\centering
  \includegraphics[width=0.59\textwidth]{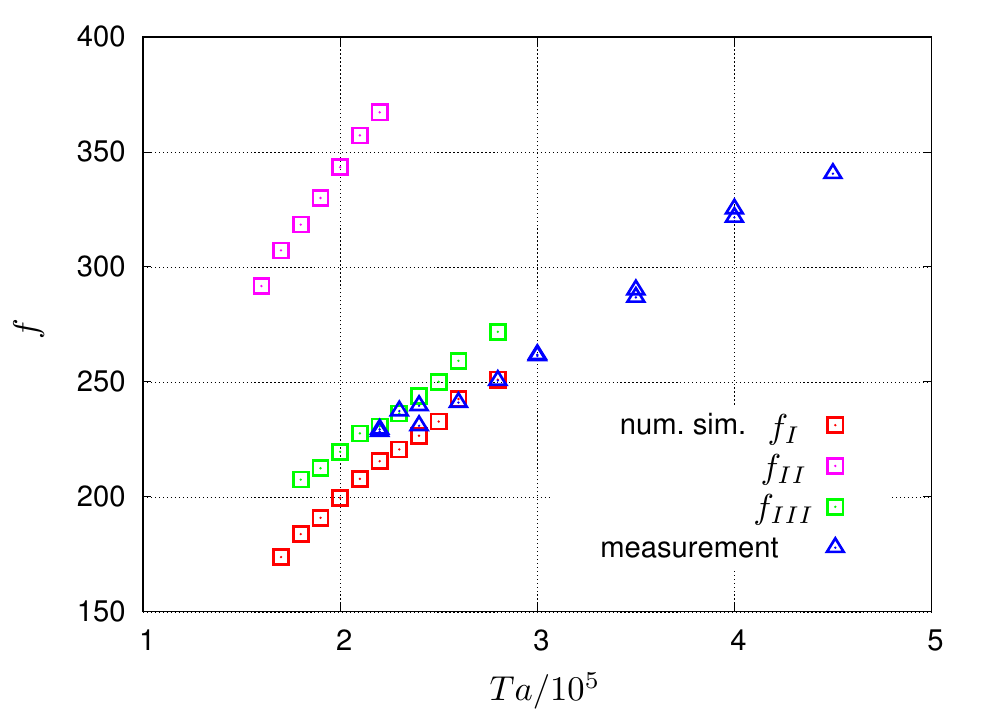}
  \vspace*{-1.2em}
\caption{\label{uxfreq} Characteristic oscillation frequencies
of the secondary flow in the plane $x=0$ as
a function of the Taylor number $Ta$ showing three different branches:
$f_I, f_{II}$ and $f_{III}$ which appears at different time intervals
during the evolution of the flow (see Fig. \ref{threefreq}).
 }
\end{figure}

In order to quantify the oscillating behaviour of the flow a discrete
Fourier analysis of the kinetic energy of the secondary flow was performed.
Fig. \ref{uxfreq} shows the characteristic oscillation frequencies 
of the secondary flow velocity as a function of the Taylor number $Ta$
in the range $Ta_1 = 1.5 \times 10^5 < Ta < Ta_2 = 4.5 \times 10^5$
comparing the experimental results and those derived from the numerical
simulation.  Here, too, a very good agreement between the measurements
and the numerical calculations can be noticed.

The frequency domain $f_{I}$ corresponds to flow oscillations, which occur
 directly after the spin-up phase.
The branch $f_{III}$ corresponds to the flow oscillations in the asymptotic
phase, which is the time period after completing the spin-up phase,
when  all growing modes have been evolved.

\begin{figure}[htb]
\centering
  \includegraphics[width=0.995\textwidth]{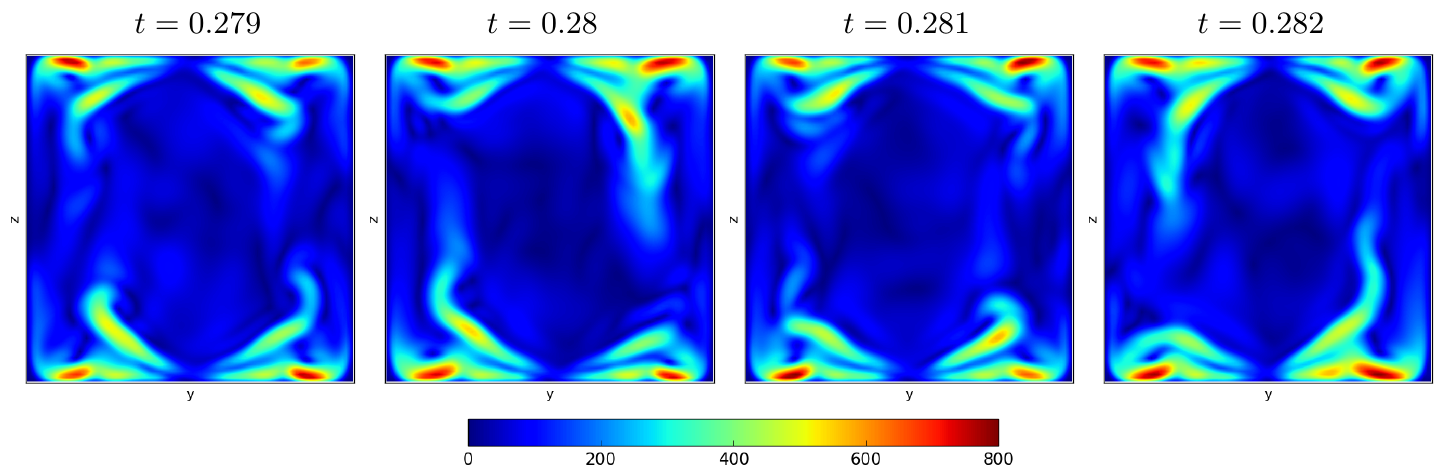}
\vspace*{-1em}
\caption{\label{snapshots_simulation}Simulation: Snapshots of the secondary flow
(contour plots of $\sqrt{(u_y^2+u_z^2)} |_{x=0}$)
for $Ta=2.6 \times 10^5$.
}
\end{figure}

\begin{figure}[htb]
\centering
  \includegraphics[width=0.995\textwidth]{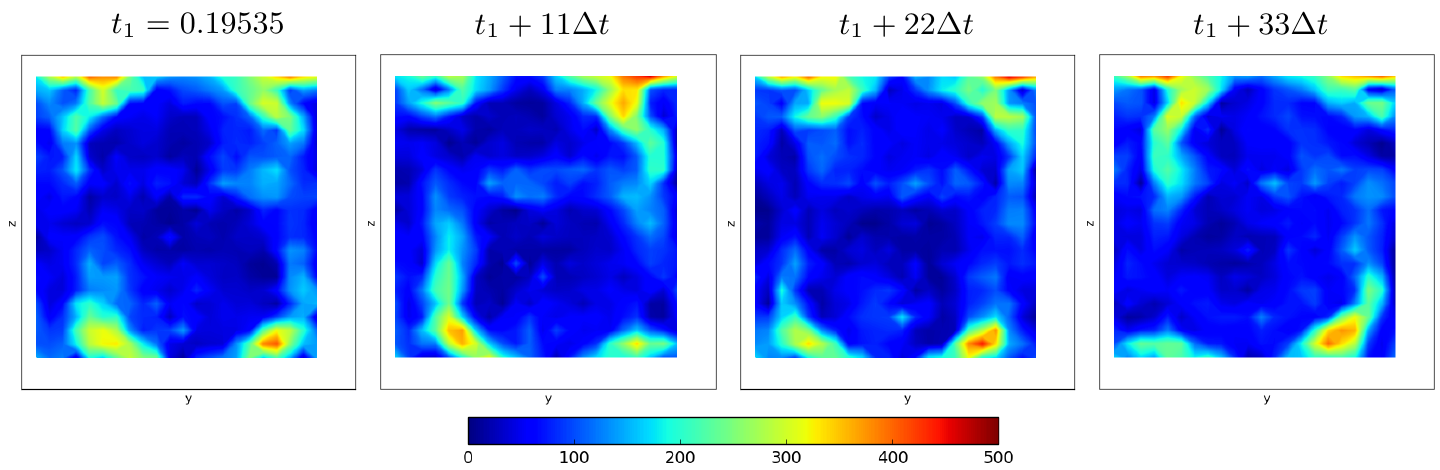} 
\vspace*{-1em}
\caption{\label{snapshots}Measured snapshots of the secondary flow
(contour plots of $\sqrt{u_y^2+u_z^2}$)
for $Ta=2.6 \times 10^5$ showing the periodic character of the flow.
The sampling interval was $\Delta t = 9.356 \times 10^{-5}$.
}
\end{figure}

Our analysis revealed that the flow structure has a periodic character
 in the range $Ta_1 = 1.5 \times 10^5 < Ta < Ta_2 = 4.5 \times 10^5$. 
The specific case $Ta = 2.6 \times 10^5$ was selected for further
examination. We applied a discrete Fourier analysis of the 
velocity data to determine the peak frequency. 
For the numerical data a value of $f_{I}=242.5$ was found
corresponding to a period of $T \approx 0.00412$.
Fig. \ref{snapshots_simulation} shows four snapshots of the secondary flow 
(contour plots of $\sqrt{(u_y^2+u_z^2)} |_{x=0}$)
for $Ta=2.6 \times 10^5$. 
The numerically predicted flow structure of the secondary flow is well 
confirmed by corresponding velocity measurements on the plane $x=0$ which
are displayed in  Fig. \ref{snapshots}. A Fourier analysis of the 
measurements resulted in a peak frequency of $f=241$ corresponding
to a period of $T \approx 4.15 \times 10^{-3}$. 
The sampling interval was $\Delta t = 9.356 \times 10^{-5}$ 
($\Delta t = \unit{0.3126}{s}$ in physical units).
We can recognize, that the flow structure has a periodic character with
the period $44 \times \Delta t = 4.12 \times 10^{-3}$, which corresponds very well
with the data coming from 
the numerical simulation for the same Taylor number 
(cf. Fig. \ref{snapshots_simulation}). 


In order to ensure mesh refinement independence,
convergence studies at different typical regimes were carried out.
For instance, at $Ta=10^5$ (steady flow regime) and at $Ta=1.9 \times 10^5$
(oscillatory flow regime showing 3 different transient intervals with 3
different oscillatory frequencies, respectively). Table \ref{table3}
gives the mean steady state secondary flow velocity for $Ta=10^5$
and the three characteristic oscillation frequencies
for $Ta=1.9 \times 10^5$ depending of the mesh refinement.

\begin{table}[hbt]
\begin{tabular}{c | cccc}
 & \hspace*{0.5em}  mesh I   \hspace*{0.05em} &
   \hspace*{0.05em} mesh II  \hspace*{0.05em} & 
   \hspace*{0.05em} mesh III \hspace*{0.05em} & 
   \hspace*{0.05em} mesh IV \\[-0.2em] \hline \hline
number of cells & $10^6$ & $1.57 \times 10^6$ & $2 \times 10^6$ & $2.46 \times 10^6$ \\ \hline
$Ta= 10^5$ & & & \\[-0.6em]
$<u_{rz}>^{1/2}$ & 107.24 & 107.39 & 107.47 & 107.51 \\ \hline
$Ta=1.9 \times 10^5$ & & & \\[-0.6em]
$f_I$     & 189.3 & 190.4 & 190.8 & 190.3 \\[-0.6em]
$f_{II}$  & 328.5 & 329.5 & 330.8 & 330.2 \\[-0.6em]
$f_{III}$ & 219.1 & 216.5  & 212.2 & 214.5 \\
\end{tabular}
\caption{\label{table3} Steady state secondary flow velocity for $Ta=10^5$ 
and characteristic oscillation frequencies at different transient flow
regimes for $Ta=1.9 \times 10^5$ depending of the mesh refinement.}
\end{table}

A characteristic of the spin-up is $t_{99}$, which denotes the time
from the onset of the magnetic field to the moment the primary flow
reaches 99 percent of the steady state velocity.
It is worth to mention, that both, $t_{spin-up}$ and $t_{99}$, are almost
identical in a broad range of the Taylor number below
the critical Taylor number $Ta_c$ but they show different scaling behaviours
with respect to the Taylor number $Ta$ (c.f. Table \ref{table2} eqn. \ref{t99}).

\begin{figure}[htb]
\centering
  \includegraphics[width=0.62\textwidth]{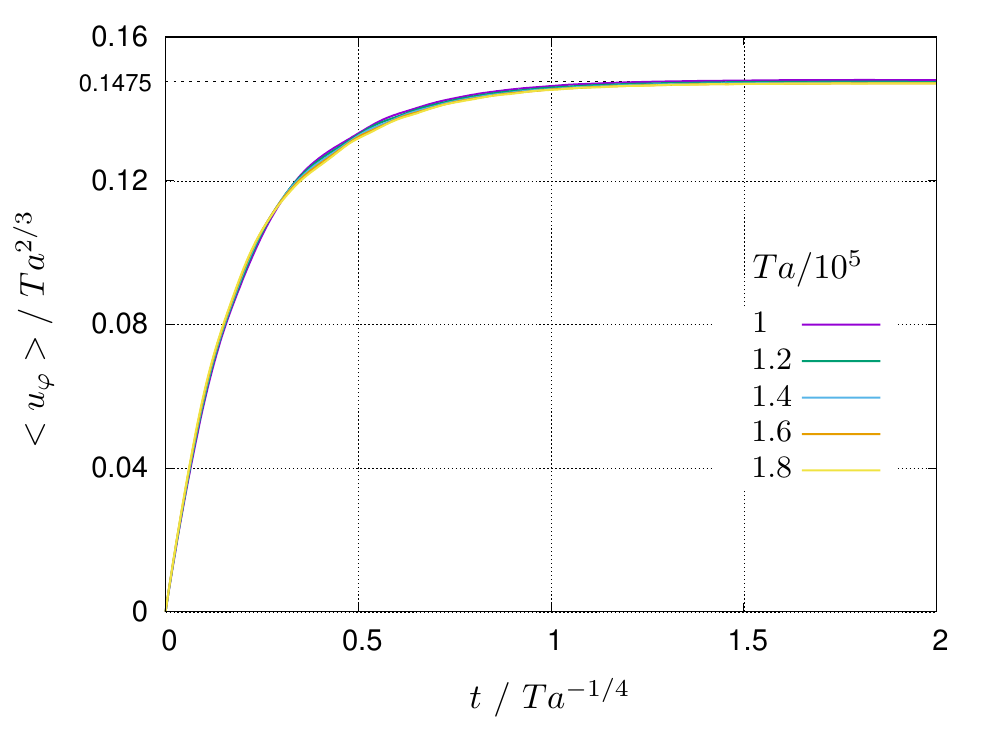}
  \vspace*{-1em}
\caption{\label{uphivst} Simulation: Time evolution of the volume
averaged  azimuthal velocity $< u_\varphi >$ using the scales
$Ta^{2/3}$ for the velocity and $Ta^{-1/4}$ for the time, respectively.
}
\end{figure}

Fig. \ref{uphivst} shows the time evolution of the volume averaged
azimuthal velocity component $<u_\varphi> = 1/V~ \int dV ~ u_\varphi$
using suitable scales.
Within this representation, the time evolution of $u_\varphi$
for different value of the Taylor number collapse to one curve.
The following scaling laws can be compiled for the interval 
$10^4 < Ta < 1.6 \times 10^5$:

\begin{equation} \label{t99}
  t_{99} \approx 1.036 ~ Ta^{-1/4} , \quad
  <u_\varphi> \approx 0.1475 ~ Ta^{2/3}  \quad \mbox{and} \quad
  \Omega \approx 0.167 ~ Ta^{3/4} \, .
\end{equation}

Here is $\Omega = (u_\varphi(x,y,0)/r)|_{r \to 0}$ the core rotation speed,
which is defined as the angular velocity at half height of the vertical center axis.

\begin{table}[hbt]
\begin{tabular}{r | cccccc}
$Ta$ \hspace*{1em}  & \hspace*{1.5em} $t_{i.a.}$ \hspace*{1em} &
  \hspace*{1em} $t_{99}$ \hspace*{1em} & $t_{spin-up}$ &
  \hspace*{0.5em}  $\D \frac{t_{99}}{t_{spin-up}}$ \hspace*{0.5em}
  & $ <u_\varphi>|_{t \to \infty}$ & $\Omega$ \\[0.75em]
\hline \hline
$10^4$          & 0.0436  & 0.1145 & 0.1168 & 1.02 & 65.86 & 155.4  \\[-0.4em]
$2 \times 10^4$ & 0.0292  & 0.0922 & 0.0927 & 1 & 108.5 & 269.8 \\[-0.4em]
$5 \times 10^4$ & 0.0174  & 0.07 &  0.0683 & 0.976 & 202.6 & 551 \\[-0.4em]
$8 \times 10^4$ & 0.0136  & 0.0615 & 0.0584 & 0.95 & 275.8 & 793.2\\[-0.4em]
$10^5$          & 0.01196 & 0.058 &  0.0542 & 0.935 &  318.9 & 942\\[-0.4em]
$1.2 \times 10^5$ & 0.01085 & 0.05575 & 0.051 & 0.915 &359 & 1082\\
\end{tabular}
\caption{\label{table2} Characteristic time scales,
mean azimuthal velocity and core rotation speed 
for different values of the Taylor number $Ta$.
}
\end{table}

\subsection{Proper Orthogonal Decomposition (POD)}
\label{section:pod}

The proper orthogonal decomposition (POD) is a powerful method
for data analysis aimed at obtaining low-dimensional approximate descriptions  
of complex flows using a model reduction. This technique provides a basis
for the modal decomposition of data recorded in the course of experiments
or numerical simulations. The POD decomposes the vector flow field into
orthogonal spatial modes and time-dependent amplitudes. For a detailed
description of POD applications in the field of computational fluid 
dynamics the reader is referred to Holmes et al. \cite{Holmes1998}.
In general, a velocity field $\bi u(\bi x,t)$ can be considered as 
the sum of a steady mean flow $\bi u_0(\bi x)$ and a fluctuating 
part $\bi u^*(\bi x,t)$:

\begin{equation}
\bi u(\bi x,t) = \bi u_0(\bi x) + \bi u^*(\bi x,t) \, .
\end{equation}

The fluctuating velocity field $\bi u^*(\bi x,t)$ is decomposed
into a sum of a limited number of proper modes

\begin{equation} \label{modedecomposition}
\bi u^*(\bi x,t) = \sum_{m=1}^{N_m} u^*_m(\bi x,t) = \sum_{m=1}^{N_m} a_m(t) \varphi_m(\bi x) \, ,
\end{equation}

where the $N_m$ functions $\varphi_m(\bi x)$ provide the orthogonal basis and 
the time dependency is represented by the respective amplitudes $a_m(t)$. Minimizing 
the projection error is equivalent to achieving an optimal description
of the kinetic energy.

In this study, we use the snapshot method which discretizes the distribution of 
the velocity variations $\bi u^*$ in space and time:

\begin{equation}
\bi u_{ij} = \bi u^*(\bi x_i,t_j) \, , \qquad \mbox{with} \quad
 i=1,...,N_x \quad \mbox{and}  \quad
 j=1,...,N_t .
\end{equation}

$\bi x_i$ are the center coordinates of the finite volumes, in which the spatial domain is sub-divided.
Now we can write the POD (Eq. \ref{modedecomposition}) in a discrete notation:

\begin{equation}
  \bi u_{ij} = \sum_{m=1}^{N_m} a_{mj} \bi \varphi_{im} \, ,
  \qquad \mbox{with} \quad a_{mj} = a_m(t_j) \quad \mbox{and} \quad
  \bi \varphi_{im} = \bi \varphi_m(\bi x_i) \, .
\end{equation}

The functions $\bi \varphi_m(\bi x)$ constitute an orthogonal basis with the inner
product defined as

\begin{equation}
  < \bi \varphi_{m} | \bi \varphi_{n} > = 
  \sum_{i=1}^{N_x}  \varphi_{im} W_i \varphi_{in} = \delta_{mn} \, ,
\end{equation}

where $W_i$ are weight functions being necessary for taking into account the 
significance of the different contributions by regions of the spatial discretization
with different volumes.
In our analysis we used the weight functions $W_i=v_i/v_{max}$ as the ratio of the
specific cell volume $v_i$ with respect to the maximum cell volume $v_{max}$.

The parallelized Python library {\tt modred} was applied for model reduction, modal 
analysis, and system identification of large systems and datasets as described 
in \cite{Belson2014}. The routine 
\mbox{\tt modred.compute\_POD\_matrices\_snap\_method} returns the modes 
$\bi \varphi_m$ and the eigenvalues $\lambda_m$ of the snapshot correlation 
matrix starting with the snapshots $\bi u_{ij}$, which can be expressed as time 
integral of the kinetic energy of mode $m$: 

\begin{equation}
\lambda_m = \int dt <u_m^*|u_m^*>  = \sum_j a_{mj}^2 \, .
\end{equation}

The modes are sorted in such a way, that 
$\lambda_1 > \lambda_2 > \lambda_3 > ... > \lambda_{N_m}$.

In order to evaluate the contribution rate of the mode $m$ to total kinetic energy
of the system, the normalized energy fraction

\begin{equation}
   E_m = \frac{\lambda_m}{\sum_{k=1}^{N_m} \lambda_k}
\end{equation}

has been used throughout the paper. \\[1em]
{\bf POD of a slightly supercritical flow} \\[1em]
At first we consider a slightly supercritical flow occurring at
$Ta=1.7 \times 10^5 > Ta_{c}$. 
To study the structure and time-dependent behaviour of the flow for this
value of the Taylor number we use the data obtained by numerical
simulations only.

We begin with the proper orthogonal decomposition of the numerical simulation
data and particularly with the data of the primary flow. i.e. the velocity
distribution on the horizontal plane $z=0$.

\begin{figure}[htb]
\begin{center}
  \includegraphics[width=0.59\textwidth]{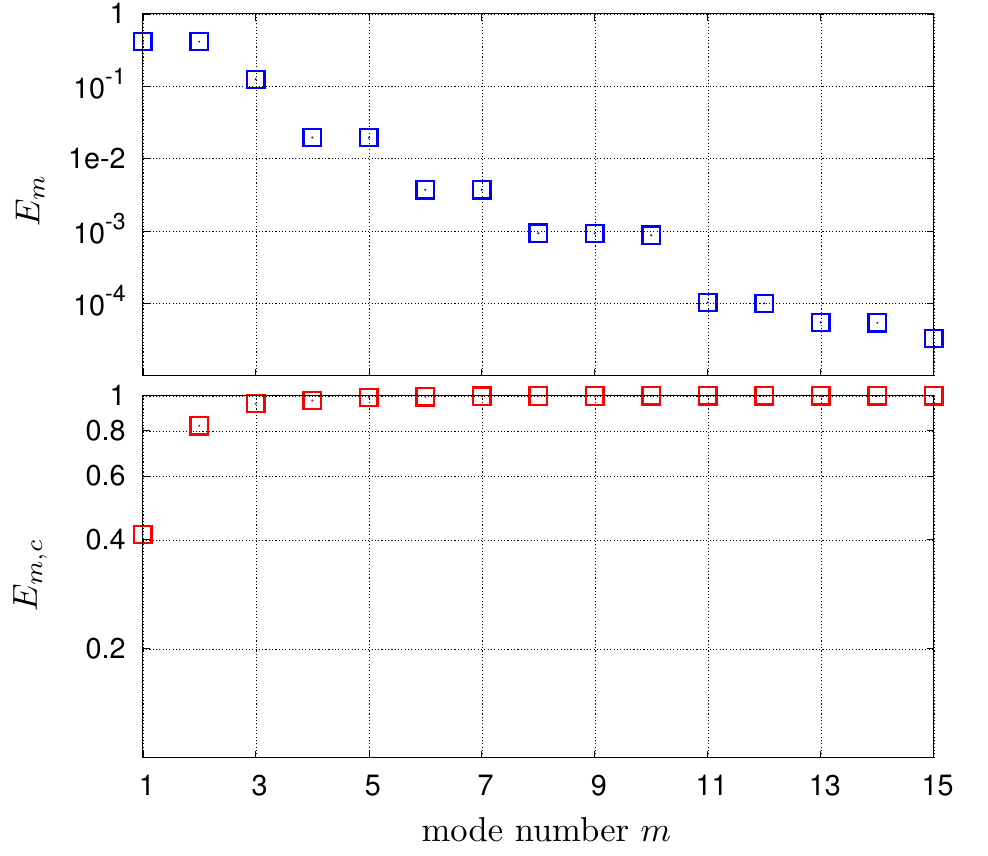}
  \vspace*{-1em}
\end{center}
\caption{\label{modeslambda} POD of the numerical simulation data of the
primary flow for $Ta = 1.7 \times 10^5$ - Contribution of each particular
mode to the total kinetic energy (top) and cumulative energy fraction
$E_{m,c} = \sum_{k=1}^m E_k$  (bottom).
}
\end{figure}

\begin{figure}[htb]
\begin{center}
  \includegraphics[width=0.78\textwidth]{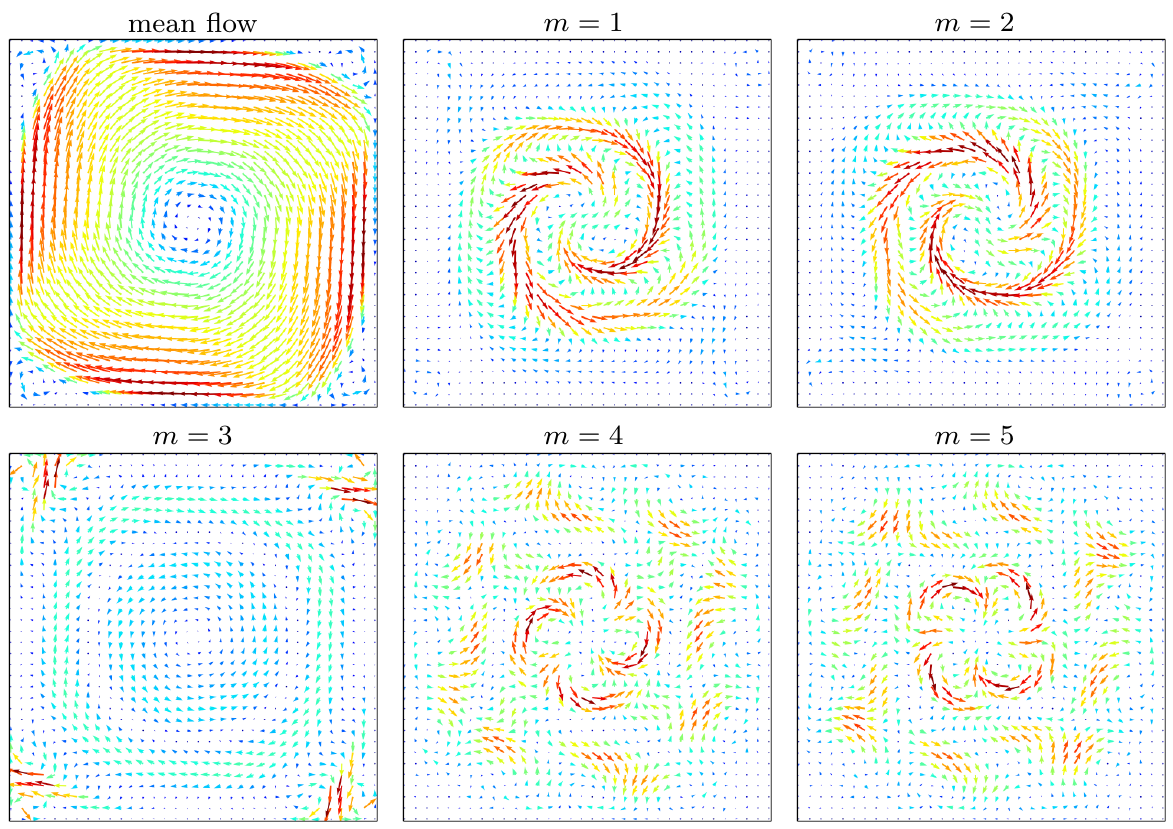}
  \vspace*{-1.1em}
\end{center}
\caption{\label{modes}Simulation: POD - Velocity vector plot of the
horizontal projection $\bi \varphi_m - \bi \varphi_m \cdot \bi e_z$
of the mode functions at the horizontal plane $z=0$ (primary flow,
$Ta = 1.7 \times 10^5$)
}
\end{figure}

The POD of the numerical data has been performed starting at the non-dimensional 
time $t=0.3$ in order to skip the spin-up phase. Fig. \ref{modeslambda} shows the 
contribution of the first 15 most important modes to the total kinetic energy. It becomes 
apparent that mode $1$ and modes $2$ have the same integral kinetic energy.
Moreover, both modes show similar flow structures (cf. Fig. \ref{modes}). 
We can see in Fig. \ref{modeslambda} that the first $5$ modes 
contain already $98.9 \%$ of the total kinetic energy of the primary flow.

\begin{figure}[htb]
\begin{center}
  \includegraphics[width=0.675\textwidth]{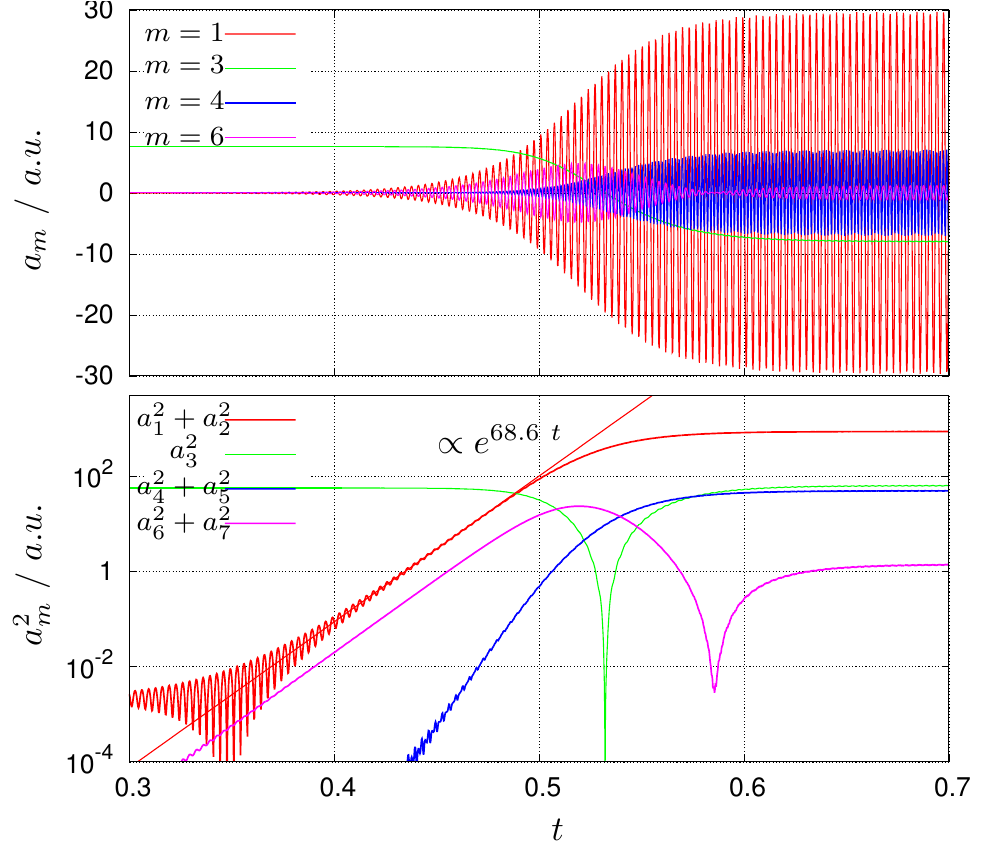}
  \vspace*{-1.5em}
\end{center}
\caption{\label{modesevolution}POD of the primary flow (numerical simulation
for $Ta=1.7\times 10^5$) - Time evolution of the amplitude of the most important 
modes $a_m(t); m=1,3,4,6$  (top) and of the kinetic energy of the modes 
$m=1,...,7$ (bottom).}
\end{figure}

Fig. \ref{modesevolution} depicts the time evolution of the amplitudes of the
leading modes $a_1(t), a_3(t) , a_4(t)$ and $a_6(t)$. The upper diagram
demonstrates the exponential growth of the modes whereas the evolution
of the kinetic energy $a_m(t)^2$ of the  modes $m=1,2,...,7$ can be seen in
the bottom graph.

We can recognize that the kinetic energy of the modes $1$ and $2$ grow
exponentially with the growth 
rate $g.r._{1,2}=68.6$. Using a discrete Fourier transform we determined 
the oscillating frequency of the amplitude $a_1(t)$ to be $f_1=306$.

Table \ref{table1} shows the growth rate of the most important POD mode of a
slightly supercritical flow for different values of the Taylor number.
An extrapolation of these values towards zero
results in a critical Taylor number of
$Ta_c \approx 1.26 \times 10^5$.

\begin{table}[hbt]
\begin{tabular}{c | p{2.1cm} p{1.8cm} p{1.8cm} p{1.8cm}}
\hspace*{0.5em} $Ta$ \hspace*{0.5em}  & \quad $1.5 \times 10^5$ & $1.6 \times 10^5$ & $1.7 \times 10^5$ & $1.8 \times 10^5$   \\ \hline \hline
$g.r.$ & \quad 11.9 & 20.5 & 34.3 & 48.6 \\
\end{tabular}
\caption{\label{table1}Growth rate of the most important POD mode of a slightly supercritical
flow for different values of the Taylor number.}
\end{table}

Fig. \ref{modes} displays vector plots of the mean flow and the horizontal 
projection of the mode functions in the horizontal plane $z=0$, i.e.
$(\bi \varphi_m - \bi \varphi_m \cdot \bi e_z)(x,y,z=0)$ (primary flow),
for the first $5$ leading modes. We can observe, that the modes $(1,2)$ 
and $(4,5)$ appear pairwise showing similar structures and the
same kinetic energy, respectively.  
While these modes can be related to differential rotation of the flow
in the horizontal cross section, mode 3 represents the transient behaviour
of the small counter-rotating vortices in the corners.

\begin{figure}[htb]
\begin{center}
  \includegraphics[width=0.95\textwidth]{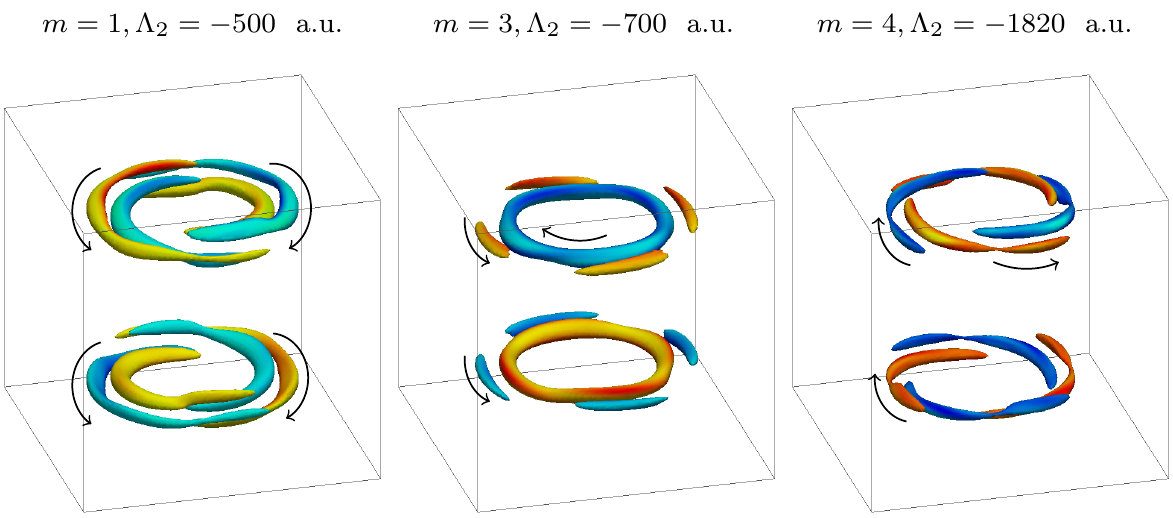} 
  \vspace*{-0.6em}
\caption{\label{lambda2}Simulation: Isosurfaces of $\Lambda_2$, the second largest
eigenvalue of the sum of the square of the symmetrical and anti-symmetrical
parts of the velocity gradient tensor for the most important modes 
$m=1$, $m=3$ and $m=4$ ($Ta = 1.7 \times 10^5$). The color indicates 
the sign of azimuthal vorticity and the arrows shows the direction of
the azimuthal velocity.
}
\end{center}
\end{figure}

For a better understanding of the three-dimensional structure 
of the principal unstable flow modes, we show in Fig \ref{lambda2}
isosurfaces of $\Lambda_2$, the second largest eigenvalue of the
sum of the square of the symmetrical and anti-symmetrical parts of 
the velocity gradient tensor for $Ta=1.7 \times 10^5$. 
The $\Lambda_2$ vortex criterion can adequately identify vortices
from a three-dimensional velocity field \cite{Jeong1995}.

The structures shown in Fig. \ref{lambda2} form pairs or quartets of
counter-rotating tubes. These vortices are responsible for the velocity
fluctuations observed in Fig. \ref{modesevolution} and for the symmetry
breaking of the basic steady flow. The red curve there corresponds to the
time evolution of the mode $m=1$ and the blue curve corresponds to 
the mode $m=4$.

In Sec. \ref{transitiontounsteadyflow}, we have seen, that the experimental data
show an oscillating behaviour, especially for the case of the secondary flow.
In analogy to the numerical simulation, we perform a
proper orthogonal decomposition of the measurement data for supercritical
flows to characterize the secondary flow.\\[1em]
{\bf POD of a supercritical flow}\\[1em]
In order to obtain a good basis for the comparison between the numerical
simulations and the experimental data, we consider in both cases the primary
and the secondary flow together as one item during the POD processing.
For $Ta < 2.2 \times 10^5$ we cannot identify in the measurements any
characteristic flow frequencies (see Fig. \ref{uxfreq}). Therefore,
the POD analysis was conducted for a Taylor number of $Ta=2.6 \times 10^5$,
which is approximately two times the critical value $Ta_{c}$.

\begin{figure}[htb]
\begin{center}
  \includegraphics[width=0.59\textwidth]{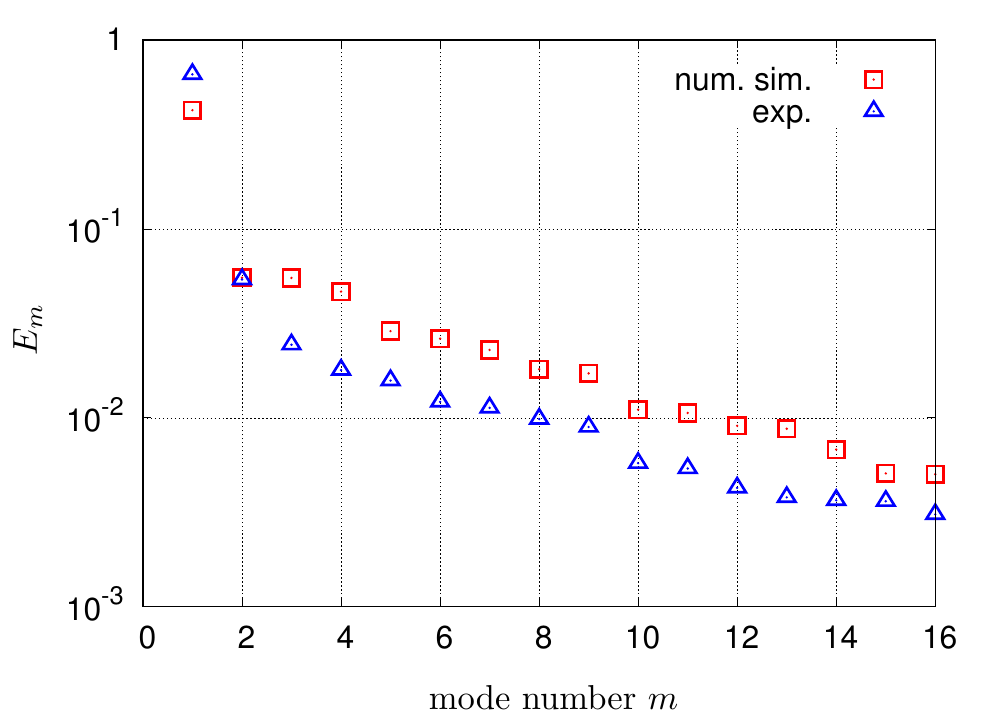}
  \vspace*{-1.1em}
\end{center}
\caption{\label{modeslambdaexp} POD of the numerical (red) and experimental data
(blue) of the secondary flow - Contribution of each particular mode to the total
kinetic energy for $Ta= 2.6 \times 10^5$
}
\end{figure}

\begin{figure}[hbt]
\begin{center}
  \includegraphics[width=0.81\textwidth]{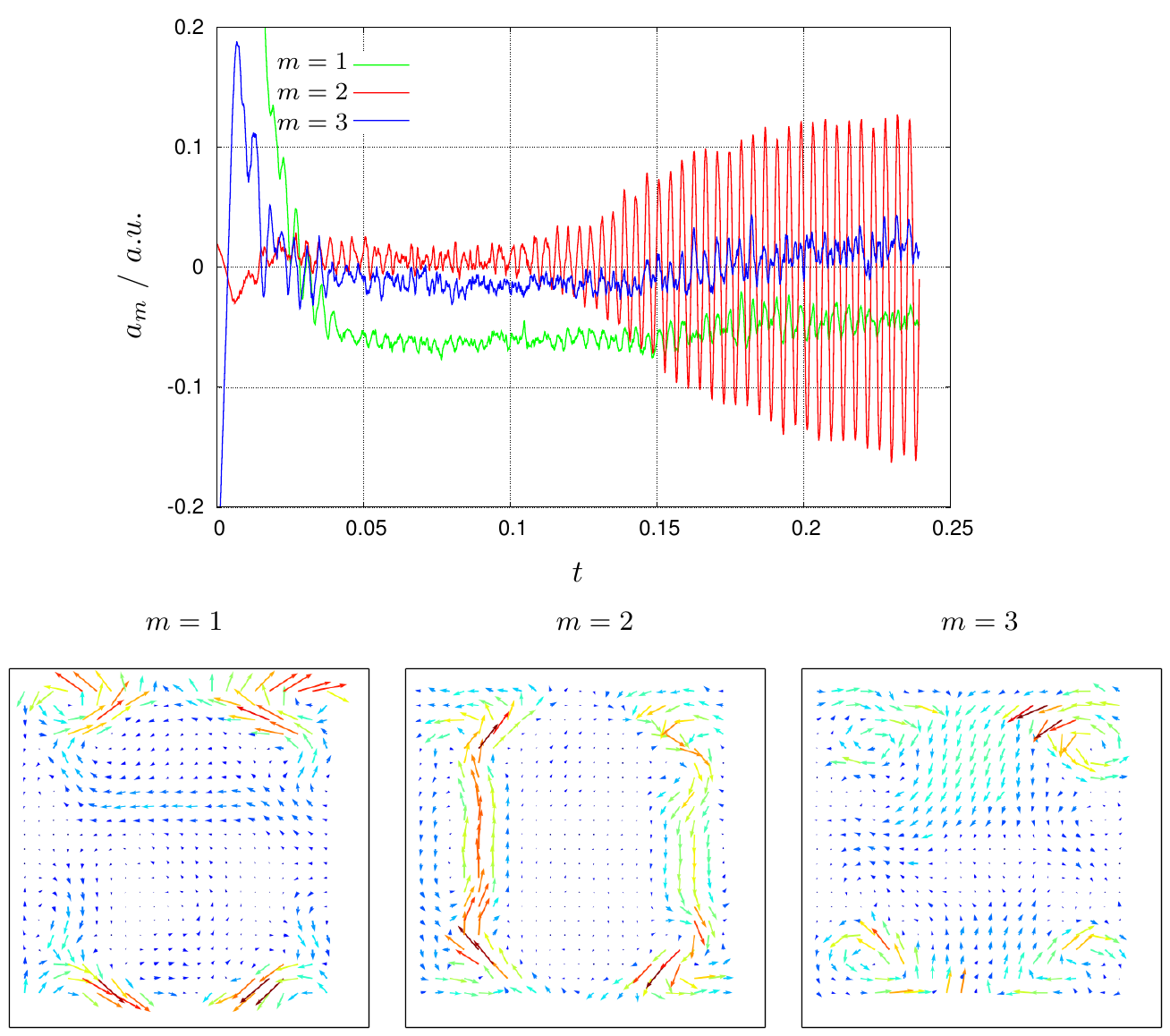}
  \vspace*{-1.3em}
\end{center}
\caption{\label{modesexp}POD of the experimental data -
Time evolution of the amplitude of the first important modes
$a_m(t)$ (top) and corresponding modes flow structures of the
secondary flow (bottom) for $Ta=2.6\times 10^5$}
\end{figure}

Fig. \ref{modeslambdaexp} shows the contribution to the kinetic energy
of each mode and Fig. \ref{modesexp} shows the
time evolution of both the amplitude of the first
important modes $a_m(t)$ (top) and the corresponding
kinetic energies (bottom) of the experimental data 
coming from the measurements of both the primary and the secondary flow 
for $Ta=2.6\times 10^5$.
The amplitude of the first important mode $m=1$ decreases during the 
spin-up phase and reveals the same structure as those for the mean flow. (see Fig. \ref{modesexp})
The amplitude of the second mode increases exponentially and
shows an oscillatory behaviour in the saturation phase for $t > 0.2$.

\begin{figure}[hbt]
\begin{center}
   \includegraphics[width=0.81\textwidth]{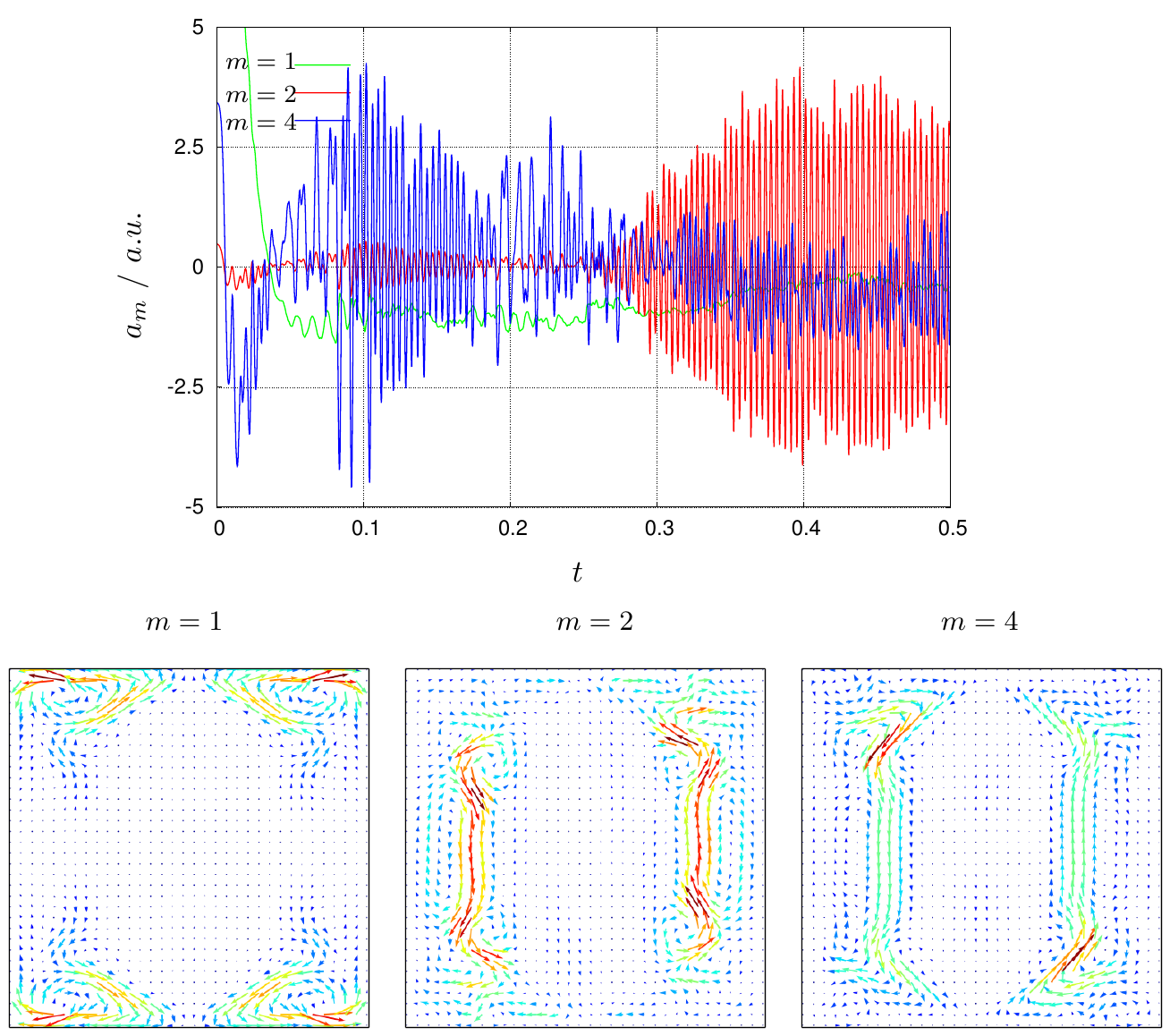}
\end{center}
\vspace*{-1.3em}
\caption{\label{modesta26}
POD of the numerical simulation - flow structure of the most important modes of the
secondary flow for $Ta=2.6\times 10^5$
}
\end{figure}

Fig. \ref{modesexp} shows the structure of the most important modes of the 
secondary flow for $Ta=2.6\times 10^5$. These results confirm very well the
corresponding findings from the numerical simulations, which are shown
in Fig. \ref{modesta26}.

Figure \ref{modesta26} shows the POD of the numerical simulation for
$Ta=2.6 \times 10^5$, which corresponds to Figs. \ref{modesexp}
coming from the experiments for the same value of the Taylor
number. We can identify that the combination of the 
modes $m=2$ and $m=3$ grows exponentially 
and shows the same structure as the experimental mode $m=2$.
Obviously, this pattern is equivalent to a recirculation roll covering the
vertical section of the cube. The instability revealed in the 
Figs. \ref{snapshots_simulation} and \ref{snapshots}
is related to this mode. 
The mode $m=1$ describes in both cases the initial spin-up phase of the flow
and has exactly the same structure as the mean flow but with opposite sign.

\section{Conclusions}
\label{conclusions}

Numerical simulations and experimental investigations were performed within
this study for investigating characteristic flow patterns arising in an
electrically conducting fluid inside a closed cubic container in consequence
of the applying a rotating magnetic field. 
Direct numerical simulations were performed using a semi-analytical expression
for the induced electromagnetic force density. 
Two-dimensional distributions of the fluid velocity in two perpendicular
planes were measured by means of a dual-plane two-component ultrasound array
Doppler velocimeter (UADV) with a high frame rate. It was demonstrated that
this instrumentation allows for reliable and accurate measurements of very
small velocities in the laminar regime ($\le \unit{1}{mm/s}$), but, can also
resolve the time-dependent flow in the turbulent region,
where the velocity can achieve values of several cm/s. 
Due to the non-deterministic onset of oscillatory instabilities,
multi-plane flow imaging with a high frame rate over long durations
is crucial to capture the flow spanning multiple time scales. 
This requirement is met by the UADV system by real-time data
compression on an FPGA and continuous streaming.
We performed UADV measurements with a frame rate up to $f=11.2~ Hz$ an
a duration of up to $5000~ s$.

Our results reveal that the fluid flow observed inside the cube shows a remarkable
resemblance to the flow structures occurring in a circular cylinder. In particular,
we found the transition from the steady state to time dependent flow structures
 at a critical $Ta$ number at $Ta_c>1.26 \times 10^5$ ($Ta_c = 1.232 \times 10^5$
for a cylinder with aspect ratio 1). With respect to the scaling behaviour of
the primary flow intensity we obtained a relationship for small values of the
Taylor number ($Ta < 2 \times 10^2$) corresponding to 
$Re_{max} \approx 0.031~ Ta$ and $Re_{max} \approx 2.08~ Ta^{1/2}$ for high 
Ta numbers ($Ta > 4 \times 10^5$), respectively. Both scaling laws assort 
well with either analytical relations or predictions
made by Davidson and Hunt \cite{davidsonhunt1987} for the case of an RMF-driven
flow in an infinite circular cylinder.

The occurrence and exponential growth of spontaneous flow instabilities was 
observed numerically and in the experiment. The characteristic frequencies of the
oscillating flow just above the critical Taylor number $Ta_{c}$ were determined.
The POD method was applied to identify the dominating modes of the flow structure.
Unlike the case of the RMF-driven flow in a circular cylinder, we did not find
Taylor - G\"ortler vortices neither in the numerical simulation nor in the 
experiments. The absence of curved walls in the cube might be the reason for
this difference. Numerical simulations and flow measurements show an excellent
agreement and provide accurate results with respect to both the mean flow 
structures and the evolution of the flow in time.

\section*{Acknowledgment}
The financial support from the German Helmholtz Association in the framework 
of the Helmholtz Alliance ``Liquid Metal Technologies (LIMTECH)''
and from the Deutsche Forschungsgemeinschaft (DFG) project BU 2241/2-1
``Ultrasonic measuring system with adaptive sound field for turbulence 
investigations in liquid metal flows'' is gratefully acknowledged.
The authors appreciate productive discussions with Dr. T. Weier
concerning the POD.

\bibliography{mybib}

\end{document}